\documentclass[preprint,12pt]{emulateapj}
\usepackage{graphicx}
\usepackage{epsfig}
\usepackage{natbib}
\usepackage{amssymb}
\usepackage{amsbsy}
\usepackage{natbib}
\usepackage{url}
\usepackage[mathcal]{euscript}
\bibpunct{(}{)}{,}{a}{}{,}

\newcommand{\rsun}{R$_{\sun}$}
\newcommand{\msun}{M$_{\sun}$}
\newcommand{\teff}{$T_{\rm eff}$}

\newcommand{\vsinis}{$v \sin i$'s}
\newcommand{\vsini}{$v \sin i$}

\newcommand{\eg}{{\it e.g.}}
\newcommand{\etal}{et~al.}
\newcommand{\ks}{$K_{\rm s}$}
\newcommand{\vmk}{$(V-K_{\rm s})_0$}
\newcommand{\vmi}{$(V-I_{\rm c})_0$}

\begin{document}
 
\def\simlt{\vcenter{\hbox{$<$}\offinterlineskip\hbox{$\sim$}}}
\def\simgt{\vcenter{\hbox{$>$}\offinterlineskip\hbox{$\sim$}}}
\def\etal{et al.\ }
\def\kms{km s$^{-1}$}

\title{Rotation in the Pleiades With K2:  III. Speculations on Origins and Evolution}
\author{John Stauffer\altaffilmark{1}, 
Luisa Rebull\altaffilmark{1}, 
Jerome Bouvier\altaffilmark{2}, 
Lynne A. Hillenbrand\altaffilmark{7},  
Andrew Collier-Cameron\altaffilmark{3},
Marc Pinsonneault\altaffilmark{4},
Suzanne Aigrain\altaffilmark{5},
David Barrado\altaffilmark{10}, 
Herve Bouy\altaffilmark{10},
David Ciardi\altaffilmark{14},
Ann Marie Cody\altaffilmark{6}, 
Trevor David\altaffilmark{7}, 
Giusi Micela\altaffilmark{9},
David Soderblom\altaffilmark{11},
Garrett Somers\altaffilmark{4},
Keivan G. Stassun\altaffilmark{13},
Jeff Valenti\altaffilmark{11},
Frederick J. Vrba\altaffilmark{12}
}
\altaffiltext{1}{Spitzer Science Center, California Institute of
Technology, Pasadena, CA 91125, USA}
\altaffiltext{2}{Universit\'e de Grenoble, Institut de Plan\'etologie 
et d'Astrophysique de Grenoble (IPAG), F-38000 Grenoble, France;
CNRS, IPAG, F-38000 Grenoble, France}
\altaffiltext{3}{SUPA, School of Physics and Astronomy, University of St. Andrews,St. Andrews,
Fife KY169SS, UK}
\altaffiltext{4}{Department of Astronomy, The Ohio State University, 140 W. 18th Avenue,
Columbus, OH 43201, USA}
\altaffiltext{5}{Department of Physics, University of Oxford, Keble Road, 
Oxford, OX1 3RH, UK}
\altaffiltext{6}{NASA Ames Research Center, Kepler Science Office, Mountain
View, CA 94035}
\altaffiltext{7}{Astronomy Department, California Institute of
Technology, Pasadena, CA 91125, USA}
\altaffiltext{9}{INAF - Osservatorio Astronomico di Palermo, Piazza 
del Parlamento 1, 90134, Palermo, Italy}
\altaffiltext{10}{Centro de Astrobiolog\'ia, Dpto. de
Astrof\'isica, INTA-CSIC, E-28692, ESAC Campus, Villanueva de
la Ca\~nada, Madrid, Spain}
\altaffiltext{11}{Space Telescope Science Institute, 3700 San Martin Drive,
Baltimore, MD 21218, USA; Center for Astrophysical Sciences, Johns Hopkins University,
3400 North Charles St., Baltimore, MD 21218, USA}
\altaffiltext{12}{U.S. Naval Observatory, Flagstaff Station, 10391 
West Naval Observatory Road, Flagstaff, AZ 86001, USA}
\altaffiltext{13}{Department of Physics and Astronomy, Vanderbilt University,
Nashville, TN  37235, USA} 
\altaffiltext{14}{NASA Exoplanet Science Institute, Caltech, Pasadena, CA, 91125, USA} 

\begin{abstract}

We use high quality K2 light curves for hundreds of stars in the
Pleiades to understand better the angular momentum evolution and
magnetic dynamos of young, low mass stars. The K2 light curves provide
not only rotational periods but also detailed information from the
shape of the phased light curve not available in previous studies.  A
slowly rotating sequence begins  at \vmk$\sim$1.1 (spectral type F5)
and ends at \vmk\ $\sim$\ 3.7  (spectral type K8), with periods rising
from $\sim$2 to $\sim$11 days  in that interval.  Fifty-two percent of
the Pleiades members in that  color interval have periods within 30\%
of a curve defining the slow  sequence; the slowly rotating fraction
decreases significantly redward  of \vmk=2.6.  Nearly all of the
slow-sequence stars show light curves  that evolve significantly on
timescales less than the K2 campaign  duration.  The majority of the
FGK Pleiades members identified as  photometric binaries are
relatively rapidly rotating, perhaps because  binarity inhibits
star-disk angular momentum loss mechanisms during  pre-main sequence
evolution.  The fully convective, late M dwarf Pleiades  members (5.0
$<$ \vmk\ $<$ 6.0) nearly always show stable light curves,  with
little spot evolution or evidence of differential rotation.   During 
PMS evolution from $\sim$3 Myr (NGC~2264 age) to $\sim$125 Myr 
(Pleiades age), stars of 0.3 \msun\ shed about half their angular
momentum, with the fractional change in period between 3 and 125 Myr being nearly
independent of mass for fully convective stars.  Our data also suggest
that very low mass  binaries form with rotation periods more similar
to each other and faster  than would be true if drawn at random from
the parent population of single stars.

\end{abstract}

\section{Introduction}

The  Kepler main mission produced multi-year light curves for
$>$100,000 low  mass stars with exquisite precision and unprecedented
cadence and duration.  Those light curves were primarily designed to
identify exoplanet transits and to determine the frequency of
terrestrial planets around  low mass host stars.  The mission was
remarkably successful, identifying hundreds of now-confirmed
exoplanets (Batalha 2014) and thousands of additional exoplanet
candidates (Rowe \etal\ 2015).

In addition to the main exoplanet science, the high quality and long
duration of the Kepler light curves enabled many other types of
science, including (amongst others) astroseismology (\eg, Aguirre
\etal\ 2014), the identification and characterization of eclipsing
binary stars (\eg, Prsa \etal\ 2011), rotation periods for low mass
field stars (\eg, McQuillan, Mazeh \& Aigrain 2014), the frequency of
super flares in solar mass stars (Maehara \etal\ 2012), the
identification of long-lived clouds in the atmosphere of an L dwarf
(Gizis \etal\ 2015), quantifying how clouds and latitudinal
differential rotation affect the optical light curves of ice-giant
planets (Simon \etal\ 2016), and AGN reverberation-mapping (\eg, Pei
\etal\ 2014). However, the fixed position in the sky of the original
Kepler field meant that a number of science topics amenable to study
with high-quality light curves were only minimally, or not at all,
addressed.  

K2, the extended Kepler mission (Howell \etal\ 2014) represents 
an unexpected opportunity to target
additional locations in the Galaxy for specific science goals.  One
particular science topic that greatly benefits from the new K2 mission
is the study of rotation and rotation-related phenomena in young, low
mass stars.  The original Kepler field contained no star-forming
regions and no young (age $\leq$ 1 Gyr) open clusters. The K2 fields
were specifically designed to provide light curves for a sample of
nearby, young pre-main sequence (PMS) stars (Upper Sco and Taurus) and
light curves for stars in several nearby, young open clusters
(Pleiades, M35, Hyades, Praesepe).  These light curves are providing a
wealth of new data important for understanding the formation and
evolution of low mass stars, and in particular their angular momentum
evolution.

The angular momentum evolution of young, low mass stars has been a topic
of great interest to stellar astronomers since at least the 1960's, when Bob
Kraft published his studies of the spectroscopic rotational velocities
(\vsini's) of A, F and early G dwarfs in the Alpha Per, Pleiades, Hyades
and Praesepe open clusters (Kraft 1965, 1966, 1967a, 1967b).  Kraft (1970)
showed that these data required that low mass stars lose angular momentum over
time, presumably due to stellar winds (Schatzman 1962; Weber \& Davis 1967).
Skumanich (1972) quantified this spindown by showing that Kraft's data 
could be approximately fit by a v$_{rot}$\ $\propto$\ t$^{-1/2}$\ relation.
With improved detectors, later spectroscopic studies of many of the same open
clusters were able to derive \vsini's down into the M dwarf range (Stauffer \&
Hartmann 1987; Stauffer, Hartmann \& Latham 1987; Soderblom \etal\ 1993b;
Terndrup \etal\ 2000); these new data highlighted that for M $<$ 1 \msun\ there is in
fact a wide range in rotational velocities on the ZAMS.  This wide range in
rotation decreases with time, such that by $\sim$600 Myr all G dwarfs have very
similar rotational velocities (Radick \etal\ 1987); the time for the rotational
velocities to converge increases with decreasing mass (Stauffer, Hartmann
\& Latham 1987).
More recently, wide-field cameras on small telescopes with large fields of view
have made it possible to obtain rotation periods for large samples of low
mass stars in many open clusters and star-forming regions, improving further 
the empirical database describing how low mass stars shed their angular momentum
over time.  These data have motivated a large body of theoretical models which
attempt with reasonable success to fit the distribution of rotation as a function
of mass and age.  Recent review articles which summarize the available rotational
data and the theoretical models include Herbst \etal\ (2007) and  Bouvier \etal\ (2014); see
also Johnstone \etal\ (2015) for a thorough discussion of the spindown models
and their uncertainties.

Another key ingredient for understanding the angular momentum evolution of
low mass stars is their magnetic field topologies.  Essentially all the angular
momentum loss mechanisms depend on the surface magnetic fields and how those
fields interact with the stellar wind or (during early pre-main sequence
evolution) the star's circumstellar disk.  It is only recently that detailed
magnetic field topologies for low mass stars have begun to become available,
primarily from Zeeman Doppler imaging (ZDI) studies (\eg, Morin \etal\ 2008;
Donati \etal\ 2008; Gregory \etal\ 2012).  Because such studies can only be
done with large telescopes and require considerable investment in telescope
time, the number of stars that have been mapped is still small and the
dependence of field topology on mass, age, and rotation rate is only beginning
to be explored.   Recent reviews of the available empirical data and inferences
on how magnetic field topologies depend on mass and age can be found in
Donati \& Landstreet (2009) and Linsky \& Scholler (2015).

The location of Field 4 on the K2 mission was chosen in order to provide
light curve data for members of the Pleiades open cluster.
Because of its proximity (136.2 pc, Melis \etal\ 2014), richness (more
than 1500 probable members,  Sarro \etal\ 2014), age (125 Myr,
Stauffer, Schulz, \& Kirkpatrick 1998) and the large body of published
ancillary data, the Pleiades offers a strikingly good match to the
capabilities of the K2 mission. Within the K2 field-of-view, $\sim$1000 
probable and possible Pleiades members could be targeted, ranging
in mass from $>$ 4 \msun\ to $<$ 0.1 \msun.  In two companion papers,
we describe the initial results of our analysis of the K2 data.
Rebull \etal\ (2016a; hereafter Paper I) determine periods for the
Pleiades members with K2 light curves and document the available
broad-band photometry for these stars.  Rebull \etal\ (2016b;
hereafter Paper II) identify stars with multiple periodic signals
and provide extensive illustration of their light curvesignatures and
interpretation.  In this paper, we examine the distribution of
rotation periods as a function of color (or, equivalently, mass) in the Pleiades and
attempt to understand better why that distribution has its observed
form.  In \S 2, we briefly review the data we have available and the
conclusions reached in Papers I and II.  In \S 3, we discuss the
slowly-rotating main sequence in the Pleiades and possible mechanisms
that may contribute to stars falling on that sequence and its shape at
Pleiades age.  We discuss the rotational velocities of Pleiades M
dwarfs in \S 4 and two sets of abnormally slowly rotating stars in \S
5.

\section{Overview of the K2 Pleiades Rotation Data }

In order to set the stage for the remainder of the paper, in this
section we briefly review the main points from Papers I and II; all of
the detailed effort to derive periods and light curve shapes, and for
assembling the ancillary data needed to interpret the rotational data
are provided in those papers.  Reviews of the previous literature
regarding rotation periods in the Pleiades and a table with cross-ids
between the K2 EPIC identification numbers and the names by which the
Pleiades members are known in the published literature are also
provided in Paper I.

Light curves for of order a thousand candidate Pleiades members were
obtained during the K2 field 4 campaign.  This represents more than
half of the known or suspected members of the cluster; most of the
members lacking K2 light curves simply fell outside the K2 FOV (see
Figure 1 of Paper I),  and therefore we believe the sample of cluster
members with light curves should be relatively unbiased.  As with any
open cluster, the membership list includes stars with a wide range of
pedigrees, ranging from essentially certain members to stars with only
one study that suggested membership. We reassessed the membership of
all the candidate stars with K2 data, as described in Paper I, and
have identified 826 stars we consider to be both good quality members
and in the brightness range where our extracted K2 light curves are
normally reliable.   Of those, 759 have one or more significant
periods from our periodogram analysis, and it is these 759 stars which
are the topic of this paper. 

\begin{figure}[ht]
\centering
\includegraphics[width=9cm]{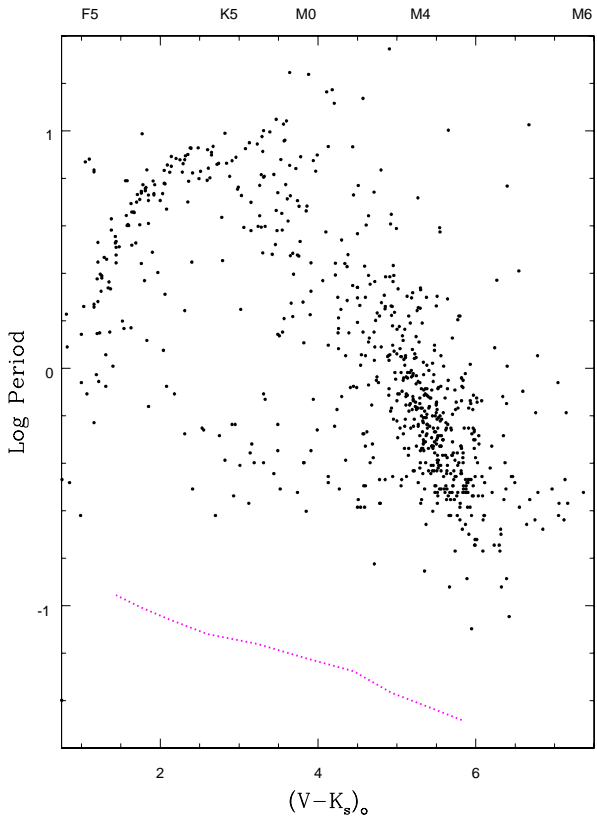}
\caption{K2 rotation periods for probable members of the Pleiades vs.\
their estimated \vmk\ colors.  The magenta
curve shows an estimate of the break-up rotation period; even the
most rapidly rotating low mass Pleiades members are more than a
factor of two away from the breakup period.
\label{fig:Figure1} }
\end{figure}

When analysing the rotation data, it is
also necessary to have some proxy for the star's mass.  
We have \vmi\ photometry for a large fraction of the FGK Pleiades
members and will sometimes uses \vmi\ as the proxy for mass in this
spectral type range
because the VI photometry is quite homogenous and the
two bands were observed nearly simultaneously, mitigating the effect
of variability on the derived color.   However, we
have chosen primarily
to use \vmk\ color as the mass proxy.  Essentially all of the Pleiades
members have 2MASS $JH$\ks\ photometry (the 2MASS 10$\sigma$ limit of
\ks= 14.5 corresponds to M6 spectral type in the Pleiades); many --
but far from all -- also have accurate, measured $V$ magnitudes. 
Where $V$ magnitudes are not known, we have used photometry at other
wavelengths to estimate \vmk\  colors (see \S 2.4 of Paper I).  
Figure~\ref{fig:Figure1} shows the main result from Paper I -- the
distribution of rotation periods for Pleiades members as a function of
their \vmk\ color.

The high quality of the K2 data makes evident that a significant
fraction (roughly a quarter) of the Pleiades members exhibit more than one
statistically significant period in their light curves.  Paper II discusses
these multiply periodic systems in great detail, and divides them into
several categories: (a) pulsational variables, most notably $\delta$\
Scuti stars (spectral type late-A or early-F; multi-periodic; periods
typically $<$ 0.1 day; normally ascribed to p-mode pulsation);  (b)
stars with two or more very different periods, probably resulting
from the target object being a binary, and each period being 
associated with one component of the binary;
(c) stars with two or more relatively
similar periods, either also ascribable to a binary system or 
possibly resulting from  latitudinal differential
rotation, spot evolution and/or spot migration; and (d) stars showing
phased light curves with two or more peaks or two or more minima, 
best explained as arising from spot groups centered at different
longitudes.  It is important to note that in Paper II, 
we have generally chosen
to utilize nomenclature for the light curve groups that are closely
linked to the empirical shapes of the light curves and the periodogram
features.  In the discussion there, we propose physical processes that
are the most likely mechanisms for generating these light curve
structures, but the primary terminology always emphasizes empirical
features.  Here, in Paper III, where the goals are to search for clues
to physical processes, we often have chosen to utilize nomenclature
which emphasizes the physical mechanisms rather than the empirical
structures.   Thus, stars whose light curves are``shape changers" in
Paper II become stars exhibiting probable spot evolution or migration
here, and FGK stars whose Lomb-Scargle (LS) periodogram show two close
peaks in Paper II become stars showing evidence of either latitudinal
differential rotation or spot evolution here.   Table 2 in Paper II provides a light
curve dictionary linking emprirical shapes to probable physical
mechanisms.  We will make considerable use of this light curve
taxonomy.

One of the key features of Figure~\ref{fig:Figure1} is the existence
of a slowly rotating sequence of stars; these are the stars that roughly
define the upper envelope to the period-color data for spectral
types F5 to late K.
This sequence becomes more prominent in older clusters
(\eg, Hyades -- Douglas \etal\ 2016;  Praesepe -- Agueros \etal\ 2011),
but has clearly begun to form at Pleiades age.  Features of the slow
sequence that we will discuss at length in the next section include
the blue edge to the sequence at \vmk $\sim$ 1.1 (\S3.2), the red edge to
the sequence at \vmk $\sim$\ 3.7 (\S3.3), and an apparent kink in the
sequence at \vmk $\sim$ 2.6 (\S3.4).

\begin{figure}[ht]
\centering
\includegraphics[width=9cm]{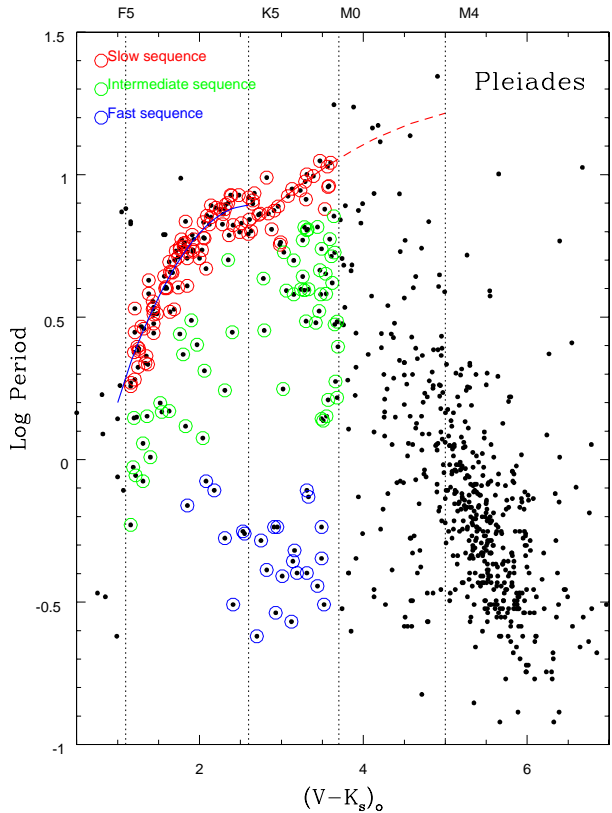}
\caption{Kepler K2 rotation periods for probable members of the
Pleiades vs.\ their estimated \vmk\ colors.  The red, green and blue
colored circles denote, respectively, stars with 
1.1 $<$ \vmk\ $<$ 3.7 having slow, intermediate and
fast rotation rates (groups which we use in our analysis in later
sections).  The left three vertical dotted lines denote the bins in \vmk\ color
used to create the histograms in Figure~\ref{fig:Figure3}.  The fourth
vertical line at \vmk\ = 5.0 corresponds approximately to the point redward
of which stars should be fully-convective.   The blue
curve is a least-squares fit to the slow sequence stars for \vmk $<$
2.6; the red curve is a fit to the slow sequence stars for 2.75 $<$
\vmk $<$ 3.7.  The red curve is extrapolated to \vmk = 5.0 in order to
roughly pass through the very slowly rotating Pleiades members beyond
\vmk\ = 3.7 (see text and \S5.2).
\label{fig:Figure2}
}
\end{figure}

We initially identified stars belonging to the slow sequence by eye.  
To define the sequence more quantitatively, we fit a polynomial curve to
those stars in the logP, \vmk\ plane.   That first fit emphasized that 
the data points seemed to show a discontinuity in period at \vmk\ $\sim$ 2.6; 
we therefore fit two separate curves to the data, one blueward and one redward
of \vmk\ = 2.6.    Because nearly all of the stars we had identified by
eye as belonging to the slow sequence had periods within $\pm$30\% of
those curves, we then used that criterion to define our final set of
slow sequence stars.  The two curves and the final set of slow sequence
stars are shown in Figure~\ref{fig:Figure2}.\footnote{The equation for 1.1
$<$ \vmk\ $<$ 2.6 is  $\log (P) = -0.8905 + 1.3658\times (V-K_{\rm
s})_0 - 0.2616\times (V-K_{\rm s})_0^2$ and for 2.6 $<$ \vmk\ $<$ 3.7
is $\log (P) = -0.1604 + 0.4809\times (V-K_{\rm s})_0 -0.0411\times
(V-K_{\rm s})_0^2$.} The extension of the red curve to \vmk = 5 is
obviously quite uncertain, but the exact shape is not important for
the purpose for which we will use it.   Finally, we have split the
remaining stars with 1.1 $<$ \vmk\ $<$ 3.7 into two other groups based
on their periods -- a set of the fastest rotators, with P(obs)/P(seq) $<$ 0.13
and a set of intermediate rotators, with 0.13 $<$ P(obs)/P(seq) $<$ 0.7,
where P(seq) is the period predicted for that \vmk\ by the polynomial
curves.   The slow, intermediate and fast rotation groups are color-coded
red, green and blue, respectively, in Figure~\ref{fig:Figure2}.  
We will utilize the polynomial curves and the three rotation 
groups in \S 3.1.   

\begin{figure}[ht]
\centering
\includegraphics[width=9cm]{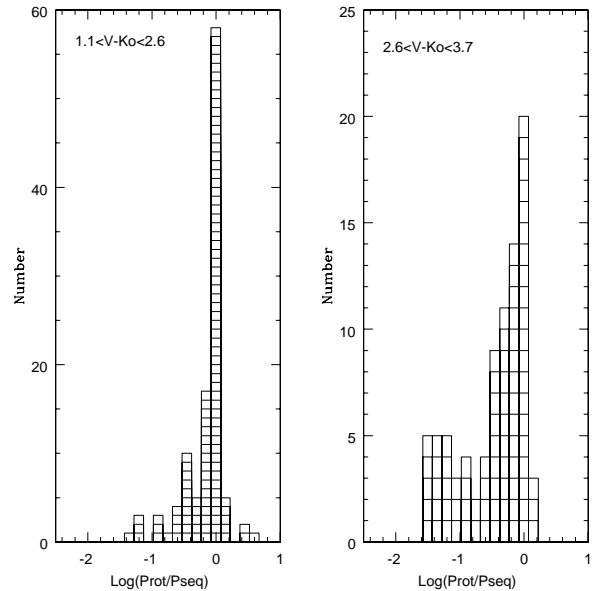}
\caption{Histograms of the Pleiades rotation period distribution for
stars with 1.1 $<$ \vmk\ $<$ 2.6 and for 2.6 $<$ \vmk\ $<$ 3.7.
The histograms were created by taking the ratio of the observed
period to the period at that color for the polynomial fits to
the slow sequence stars.  
Two thirds of the stars in the bluer interval (left panel)
have periods within 30\% of the polynomial fit versus only one third
for the stars in the redder color interval (right panel).
\label{fig:Figure3}
}
\end{figure}

For the G and K dwarfs in clusters older than about 600 Myr, all (or
nearly all) stars are on the slow sequence (Stauffer, Hartmann, \&
Latham 1987; Radick \etal\ 1987; Agueros \etal\ 2011; Delorme \etal\
2011).   By contrast, 1-3 Myr old PMS stars of this mass show a wide
range of rotational velocities  (\eg, Herbst \etal\ 2001; Carpenter
\etal\ 2001; Rebull \etal\ 2002a;  Rodriguez-Ledesma \etal\ 2009;
Affer \etal\ 2013). The collapse from the wide initial range in
angular momenta to a nearly unimodal distribution at a given mass by
Hyades age is normally attributed to a disk-locking mechanism that
temporarily inhibits spin-up for PMS stars with actively accreting
disks (Ghosh \& Lamb 1979; K\"onigl 1991; Cameron, Campbell \&
Quaintrell 1995; Bouvier \etal\ 2007) and to
stellar winds and an angular momentum loss rate that is largest for
rapidly rotating stars (\eg, Skumanich 1972; Stauffer \& Hartmann
1987; Soderblom \etal\ 1993a; Collier Cameron \& Jianke 1994;  Sills
\etal\ 2000; Gallet \& Bouvier 2013, 2015).  The distribution of rotation
periods as a function of mass for a cluster where the G and K dwarfs
have just arrived or are near to the Zero Age Main Sequence (ZAMS) --
as is the case for the Pleiades -- therefore becomes an important
fiducial dataset with which to confront angular momentum evolutionary
models.   We provide one representation of this distribution here by
taking the  rotation periods in Figure~\ref{fig:Figure2}, dividing those
periods by the period inferred from
the  polynomial fits to the slow sequence, 
and determining the relative period distribution
as a function of \vmk\ color.  These distributions are plotted as
histograms in Figure~\ref{fig:Figure3}, for the color bins indicated
by the vertical dashed lines in Figure~\ref{fig:Figure2}. 
The fraction of
stars included in the slowly-rotating sequence decreases to lower mass
(redder color).  Specifically, if one takes $N$(slow) to be the number
of stars within 30\% of the value predicted by the polynomial curves
shown in  Figure~\ref{fig:Figure2} (corresponding to the
stars circled in red in that figure) and $N$(tot) to be the total number
of stars with periods in that color bin, the ratio of $N$(slow) to
$N$(tot) is 0.69, 0.32, and 0.02 for the color intervals
1.1 $<$ \vmk\ $<$ 2.6, 2.6 $<$ \vmk\ $<$ 3.7, and 3.7 $<$ \vmk\ 5.0,
respectively.

\section{The Rotation Patterns of 0.6 to 1.1 \msun\ Stars at Pleiades Age}

We split our discussion of the Pleiades K2 data into two main color (or
mass) groupings.  Here, we discuss stars with 1.1 $<$ \vmk\ $<$ 3.6,
roughly corresponding to a mass range of 0.6 to 1.1 \msun.  In \S4,
we discuss M dwarfs, corresponding to masses $<$ 0.6 \msun.

\subsection{Why are Some Pleiades FGK Stars so Slow?}

\begin{figure}[ht]
\centering
\includegraphics[width=9cm]{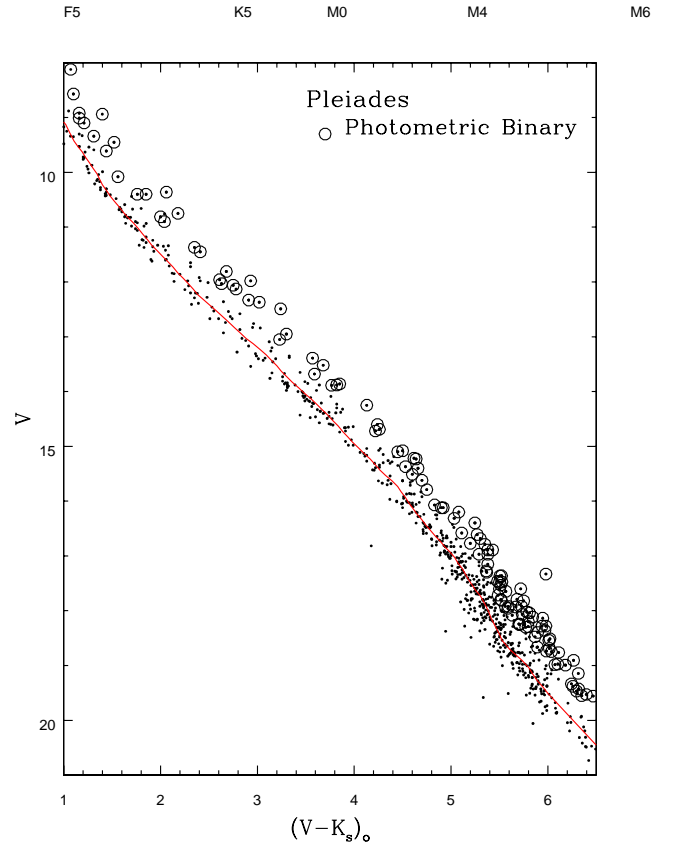}
\caption{CMD showing the Pleiades members for which we have derived
rotation periods, marking the stars identified by us as photometric
binaries with circles.
\label{fig:Figure4}
}
\end{figure}

At constant mass, the  K dwarfs in the Pleiades show a range of
$\sim$40 in their observed rotation periods (i.e. from $\sim$0.25 days
to $\sim$10 days), and hence a similar range in their
inferred total angular momentum assuming solid-body rotation.   
Understanding why this very large
dispersion in rotational histories exists has obvious importance for
understanding both star and planet formation.  
Theoretical models for angular momentum evolution (\eg, Gallet \&
Bouvier 2013, 2015) can account for this large range in rotation at a given
mass by adjusting parameters such as the disk-locking timescale, 
core-envelope coupling timescale and wind saturation velocities.
However, it is possible that other mechanisms may also contribute.
Binarity is one obvious
potential mechanism for affecting the angular momenta of ZAMS age
stars because binary companions have at least the
potential to truncate circumstellar disks and reduce the efficacy of
PMS disk-locking (Patience \etal\ 2002; Meibom \etal\ 2007).   In
order to test this idea, we have identified a sample of photometric
binaries among the K2 Pleiades sample using the $V$ vs.\ $V-I_{\rm C}$
and  $V$ vs.\ \vmk\ color-magnitude diagrams (CMDs). In the former
(which uses photometry from Stauffer \etal\ 1987 and references
therein, and from Kamai \etal\ 2014), we identify stars as photometric
binaries if they lie more than 0.3 mag above a single-star main
sequence locus.  For the latter, using the estimates of \vmk\ as
described in Paper I, we adopt $\Delta V$ = 0.45 mag for \vmk $<$ 5,
and $\Delta V$ = 0.55 mag for \vmk\ $>$ 5 to identify probable
photometric binaries.  The required $\Delta$V is larger for the \vmk\
CMD because of the lower accuracy of the (less homogeneous) \vmk\
data; the limit changes at \vmk = 5 because the slope of the
single-star locus increases at about that color.  Stars selected as
binaries from the \vmk\ CMD are illustrated in
Figure~\ref{fig:Figure4}.  The advantage of this purely photometric
selection technique is that we can apply it for every star for which
we have K2 rotation period data.

\begin{figure}[ht]
\centering
\includegraphics[width=9cm]{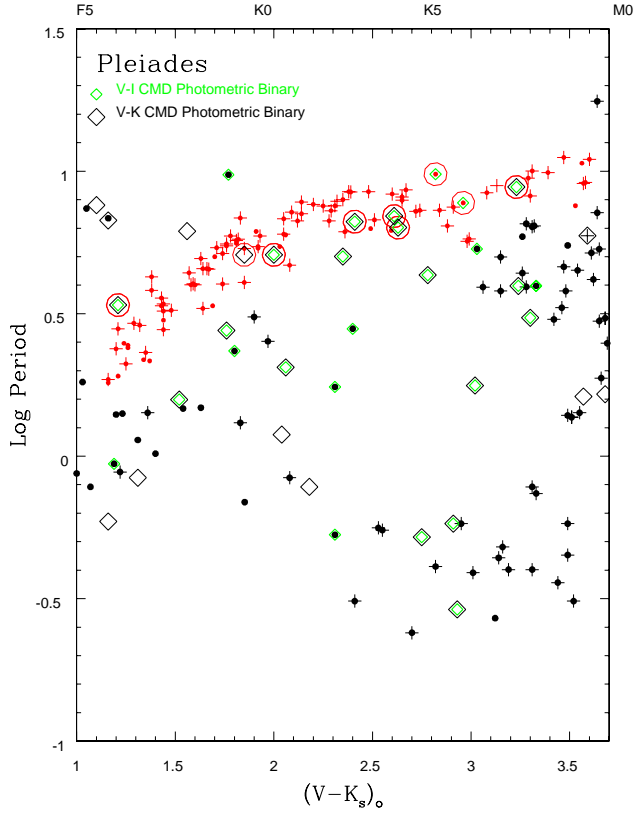}
\caption{Period-color plot, marking photometric binaries.   Dots and
plus signs are photometric singles, based on the \vmk\ and \vmi\ CMD's,
respectively.  They are colored red if on the slow sequence, and black
otherwise.  Photometric binaries are shown as diamonds, in green if
based on the \vmi\ CMD and in black if based on the \vmk\ CMD.  
Photometric binaries that are on the slow sequence are additionally 
marked by large red circles.
Given the differing sensitivities of the two CMD's to binaries,
the sorting into single and binary classes is typically
quite consistent.  Members of the slowly rotating sequence are 
seldomly photometric binaries. \label{fig:Figure5}
}
\end{figure}

We replot the period-color data for the Pleiades FGK stars (roughly
stars with 1.1 $<$ \vmk\ $<$ 3.7) in Figure~\ref{fig:Figure5}, where
the photometric binaries we have identified are now shown as
diamonds.   This diagram  shows that
members of the slow sequence are infrequently identified as
photometric binaries (9 photometric binaries out of 105 total), 
whereas the stars with intermediate rotation
periods are photometric binaries with significantly higher frequency.
In Paper I (\S 3.2, Figure 11), we show that more than a dozen stars in 
the slowly rotating sequence have been identified as spectroscopic 
binaries.  Nearly all are single-line spectroscopic binaries, 
consistent with many of them not having been
identified by us as photometric binaries.  Our association of being
a photometric binary and having somewhat more rapid rotation could be
reconciled with these additional spectroscopic binaries in the slow
sequence if the physical mechanism leading to relatively rapid rotation
operates most efficiently for nearly equal mass systems having relatively
large semi-major axes.

\begin{figure}[ht]
\centering
\includegraphics[width=9cm]{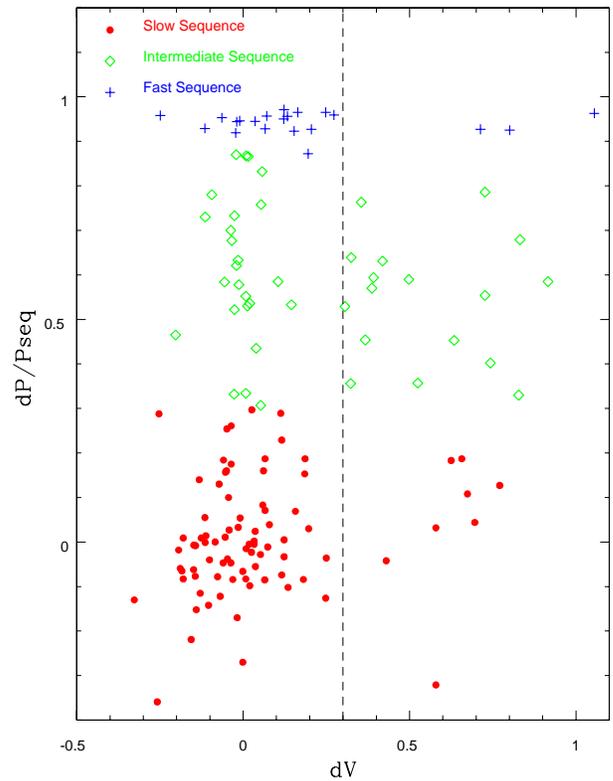}
\caption{Correlation between the displacement of stars above the main
sequence locus curve in the $V$ vs.\ $V-I_{\rm C}$ CMD ($dV$) with a
measure of how much the star's rotation period differs from the slow 
sequence locus curves.  The $y$-axis is $dP/P_{\rm seq} =
(P_{\rm seq} - P_{\rm rot})/P_{\rm seq}$, where P$_{\rm seq}$ is the
period at that color for the curves shown in Figure~\ref{fig:Figure2}.
Stars more rapidly rotating than the slow sequence locus have positive
values of $dP/P_{\rm seq}$; stars displaced above the single star locus have
positive values of $dV$.  The plot demonstrates that a large
fraction of the stars with intermediate rotational velocities are
likely photometric binaries.  One star, HII 303, has $\Delta$V $>$ 1.1
and does not appear; it is a member of the intermediate rotation group
and has been identified as at least a triple system (Bouvier \etal\ 1997).
HII 3197, the fast rotator at $\Delta$V = 1.05 was also identified as
a triple by Bouvier \etal\ 1997.
\label{fig:Figure6}
}
\end{figure}

In order to illustrate this same point in a more quantitative way, 
we have calculated ``vertical"
displacements of each star from the single-star $V$ vs.\ \vmi\ locus
($dV$) and normalized period displacements from the slow sequence locus
curves ($dP/P_{\rm seq}$) -- where $P_{\rm seq}$ is the period predicted
by the polynomial fits --  and  plot those two indices in
Figure~\ref{fig:Figure6}.  We use \vmi\ rather than \vmk\ for this
purpose because the \vmi\ photometry is more accurate, it is
less affected by spottedness (Stauffer \etal\ 2003), and in
this magnitude range we have published \vmi\ photometry for more than
three quarters of the Pleiades members (Stauffer \etal\ 1987 and
references therein; Kamai \etal\ 2014).  Figure~\ref{fig:Figure6} shows
that the stars with intermediate rotation
rates are much more frequently displaced significantly above the
single-star CMD locus than either of the other two groups.  The
fraction of stars more than 0.3 mag above the single star locus is
10\% for the slow-sequence, 41\% for the intermediate-sequence, and
14\% for the fast sequence.   Collapsing the distribution in the other
axis, 62\% (18 of 29) of the stars identified as photometric binaries 
vs.\ 22\% (26/120) of the single stars are members of the intermediate
rotation group.   Just from binomial statistics, the probability
that so many of the intermediate rotation group are photometric
binaries is less than one part in a thousand, while the probability
that so few of the slowly rotating group are photometric binaries is
of order one in a hundred.  We therefore conclude that there is very
likely a real link between rotation rate at Pleiades age in this mass
range and having a comparable mass companion.  
A plausible but not unique physical explanation for
this correlation is that binary stars with some range of parameters
(mass ratio and separation presumably being most important) inhibit
disk-dependent PMS angular momentum regulation mechanisms.\footnote{This 
tendency for young GK binaries to be more rapidly
rotating than their isolated counterparts could possibly explain a
long-standing mystery in the Hyades.   Pye \etal\ (1994) showed that
the K dwarf binaries in the Hyades have, on average, much larger X-ray
luminosities than the single K dwarfs in the Hyades.  Those binaries
have long orbital periods, so tidal spinup is not an  answer.  If our
Pleiades conjecture is correct, the Hyades K dwarf binaries could be
more rapidly rotating than their single counterparts, thereby leading
to their excess X-ray luminosities.}

\begin{figure}[ht]
\centering
\includegraphics[width=9cm]{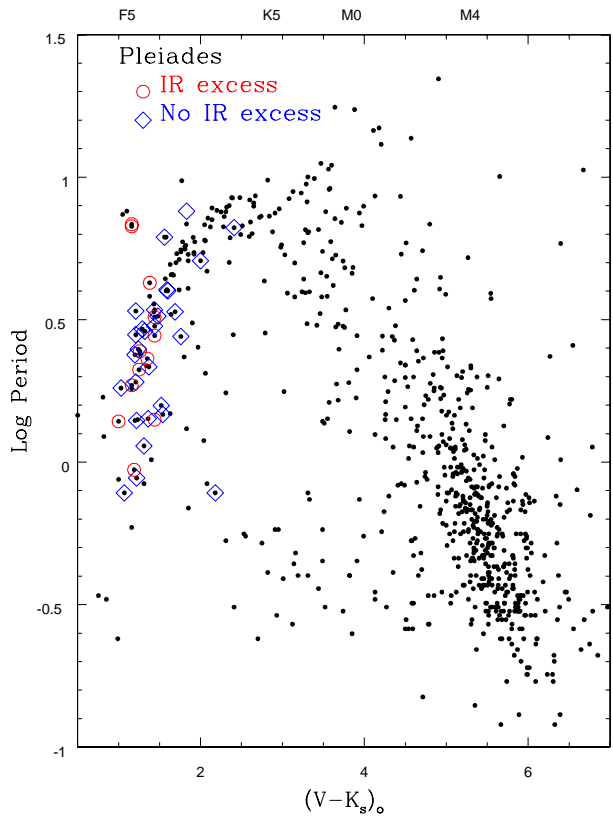}
\caption{Period-color plot, marking stars with debris disks from
Sierchio \etal (2010). In the color range where there are good Spitzer
24 $\mu$m data,  essentially all the stars are on or near the slowly
rotating sequence. Determining if debris disks and stellar rotation
correlate at Pleiades age requires new IR data for lower mass stars.
\label{fig:Figure7}
}
\end{figure}

If the correlation between binarity and rotation rate arises due to
the effect of binarity on the lifetime or amount of dust in the inner
disk during PMS evolution, 
it is possible that this effect could be reflected in the
presence or strength of debris disk signatures at later epochs.   A
number of models predict that the rotation period of low mass  PMS
stars will be locked to the Keplerian rotation period of their inner
circumstellar disk (K\"onigl 1991; Bouvier \etal\ 2007).   This
suggests the possibility that the stars with the longest-lived PMS
circumstellar disks might arrive on the main sequence as slow rotators
(\eg, Bouvier \etal\ 1997).   If those long-lived disks generate
more planets and planetessimals, then it is possible that their
signature at later epochs could be more dust from planetessimal
collisions which would be detected in the mid to far infrared as
debris disks.  We might then expect a correlation where
the stars in the slowly-rotating sequence  are also frequently
identified as having detectable debris disks.  Such a  correlation was
noted by Sierchio \etal\ (2010) but the statistical significance was
meager, in part because only $v \sin i$ data were available for many
of the stars at that time.   In Figure~\ref{fig:Figure7}, we again
replot the K2 period-color data for the Pleiades, this time
highlighting stars with detected debris disks as red circles. 
Unfortunately, in the color range where there are detections
(basically spectral type F and early G), there is no clear separation
between a slowly and rapidly rotating sequence, and therefore (in
agreement with Sierchio \etal) we cannot draw any conclusions
concerning a possible correlation of debris disks and rotation.  The
stars of interest are too faint for WISE; a test of the hypothesized
correlation between debris disk IR excess and rotation
must await a more
sensitive set of far IR data, extending into the late G and K spectral
types in the Pleiades.

\subsection{The Blue Edge - At What Color Do Non-Axisymmetrically Distributed Spots First Appear?}

High resolution spectra obtained more than 50 years ago (Anderson
\etal\ 1966) showed that, on average, rotational velocities for
Pleiades members more massive than the Sun decrease rapidly going from
early to late F.  Improved \vsinis\ for Pleiades F stars were later
obtained by Soderblom \etal\ (1993b) and by Queloz \etal\ (1998). 
Those data showed mean rotation rates decreasing from about 50 \kms\
at F2 to about 15 \kms\ at G0.  Therefore, the most rapidly rotating
Pleiades F dwarfs are at the earliest subtypes.  However, because
early F dwarfs  are not expected to have outer convective envelopes,
they should not have magnetic dynamos nor solar-like star spots.  
Without spots, these stars should not have rotationally-driven
photometric variability, and hence they should not appear in our K2
period catalog.  Outer-convective envelopes of sufficient depth to
generate significant  energy from a magnetic dynamo are believed to
first appear around spectral type F5 (Wilson 1966), so it is near that
spectral type where one might expect to identify spot-induced
rotational variability.   Our K2 data should allow us to accurately
determine  that transition point.   Unfortunately, ZAMS F stars also
occupy a \teff\ range where pulsation is expected to generate periodic
photometric variability  (Dziembowski 1977; Dupret \etal\ 2004).  In
the late-A to early-F spectral range, one expects to find $\delta$ Sct
pulsators; for somewhat cooler stars -- spectral types generally about
F0 to F3 -- one expects $\gamma$ Dor type pulsation (Balona \etal\
2011).    From the perspective of this paper, it is fortunate that the
$\delta$ Sct's have typical periods of 0.03-0.3 days (Balona \etal\
2015), allowing their pulsational variability to be easily
distinguished from the rotational modulation of F stars where one
expects periods greater than 0.5 day (also see Paper II).

\begin{figure}[ht]
\centering
\includegraphics[width=9cm]{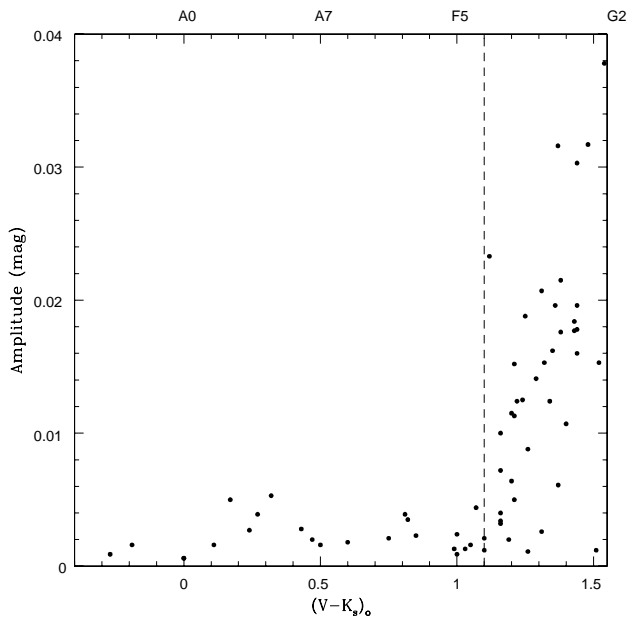}
\caption{Light curve amplitude, as defined in Paper I,
versus \vmk\ color for the A and F stars in the Pleiades where we have
identified at least one significant period in the Lomb-Scargle
periodogram.  The vertical dashed line marks the boundary we have adopted
as the blue-edge of the region where rotational modulation by magnetic
spots is assumed to drive the observed photometric variability in our
K2 Pleiades sample. See the text for a discussion.
\label{fig:Figure8}
}
\end{figure}

$\gamma$\ Dor variables, on the other hand, are expected to have
dominant periods of order 0.4 to 3.0 days (Kaye \etal\ 1999),  placing
their pulsation range directly in the range of rotation periods
expected for ZAMS age F dwarfs.   Indeed, while Balona \etal\ (2011)
identify several hundred stars with light curves in the main Kepler
field as probable $\gamma$\ Dor variables, they also acknowledge that
the variability of many of those stars may in fact be due to
rotational modulation and star spots.  For our purposes, we choose to
avoid this issue by relying on the amplitude of variability (as
determined in Paper I) to define the boundary redward of
which we are reasonably certain that spots dominate the variability
(and the periods we find are the star's rotation period).  
Figure~\ref{fig:Figure8} shows the variation of the amplitude of
photometric variability for stars where we determine periods vs.\
\vmk\ color.  There is a fairly sharp boundary at \vmk $\sim$ 1.1,
redward of which the amplitudes become much larger; the spectral type
corresponding to that  color (from Pecaut \& Mamajek 2013, Table
5) is F5, the same boundary where one begins to see significant
chromospheric emission.   Immediately blueward of that boundary, we believe it is
entirely possible that much of the variability seen in our K2 light
curves is rotational modulation due to small (possibly transient)
magnetic spots, but separating that variability from pulsational
signatures is not simple and should be the subject of a dedicated
paper.  From analysis of Q3 Kepler data for active stars,
Reinhold \etal\ (2013) also concluded that spectral type F5 marked a
dividing line, blueward of which it was much more difficult to
separate rotational variability from pulsational variability.\footnote{As shown later in Figure~\ref{fig:Figure9}, \vmk\ = 1.1 is
also approximately the color where the slow sequence locus curve for
Praesepe (age $\sim$ 600-800 Myr, Brandt \& Huang 2015) 
overlaps with the Pleiades slow locus
curve.  That intersection presumably marks the point where main
sequence angular momentum loss rates become negligible, further
reinforcing the conclusion that this is where convection-driven
magnetic dynamos become important.}

\subsection{The Red Edge of the Slowly Rotating Sequence}

There is a clearly defined slow-rotation sequence in our K2 Pleiades
data, running from \vmk $\sim$ 1.1 or spectral type F5 (see discussion
in the preceding section) all the way to \vmk $\sim$ 3.7, spectral type
K8.   The mean rotation period increases from  $P \sim$ 2 days at the
blue edge, to $P \sim$ 11 days at the red edge.  Is the 
relatively sharp cutoff to the slow sequence at \vmk $\sim$\ 3.7 simply
a direct and readily measureable indicator of the cluster's age, or is
there some other physical mechanism that contributes to this sharp end
to the slowly rotating sequence at this particular age? We suspect the
former answer is mostly correct and we provide one figure which
strongly supports this view; however, we also offer one (possibly
coincidental) piece of evidence in favor of the latter idea.

\begin{figure}[ht]
\centering
\includegraphics[width=9cm]{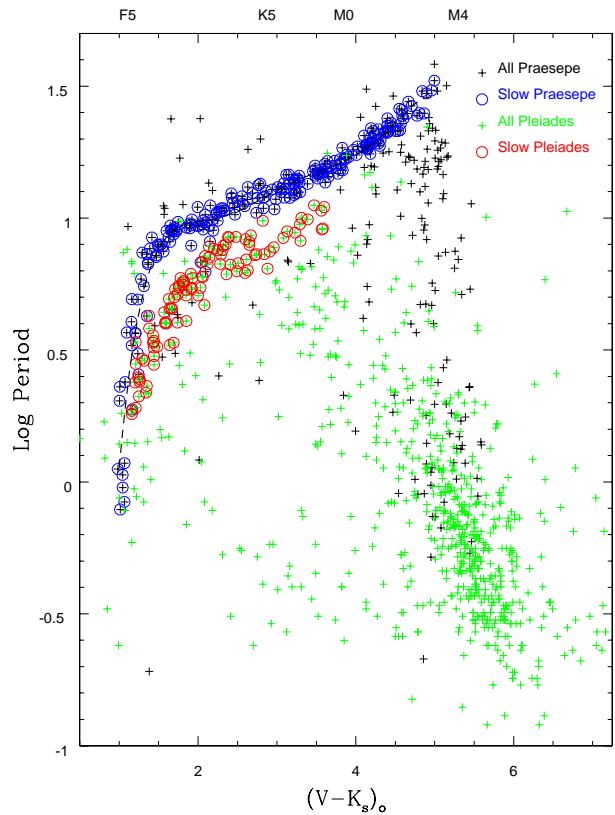}
\caption{Kepler K2 rotation periods for probable members of the
Pleiades compared to K2 rotation periods for Praesepe.  The slow
sequence evolves to longer period -- and extends much redder -- by
Praesepe age ($\sim$600 -- 800 Myr), compared to the Pleiades  sequence. 
\label{fig:Figure9}
}
\end{figure}

Direct evidence that the location of the red edge for the slowly
rotating sequence is a signpost of the age for a coeval population of
stars comes from comparison of the Pleiades K2 rotation period data to
similarly processed and analysed K2, Field 5 data for the much
older ($\sim$600 -- 800 Myr) Praesepe cluster.   We have conducted a quick-look analysis of
the publicly available K2 Field 5 light curves for
Praesepe for the sole purpose of making the comparison to the Pleiades
data.   Figure~\ref{fig:Figure9} overplots the period data for
Praesepe versus that for the Pleiades; the solid-curve is a polynomial
fit to the well-defined Praesepe slow sequence.\footnote{The
equation for the polynomial fit to the Praesepe slow sequence is:
$\log P = -13.190 + 30.079\times (V-K_{\rm s})_0 - 25.9416\times
(V-K_{\rm s})_0^2 + 11.59469 \times (V-K_{\rm s})_0^3 - 2.82725 \times
(V-K_{\rm s})_0^4 + 0.35736 \times (V-K_{\rm s})_0^5  - 0.018320
\times (V-K_{\rm s})_0^6. $}  Looking at the slow sequences
for the two clusters near \vmk $\sim$ 3.5, it is obvious that the slow
sequence extends much further to the red at $\sim$700 Myr than at 125 Myr
and therefore the location of this edge is at least partially a
measure of the age of the cluster.   A well-defined slow sequence
at 500-800 Myr extending to early M is also quite evident in the
ground-based rotation studies of Praesepe (Agueros \etal\ 2011) as
well as M37 (Hartman \etal\ 2009) and the Hyades (Delorme \etal\ 2011).

\begin{figure}[ht]
\centering
\includegraphics[width=9cm]{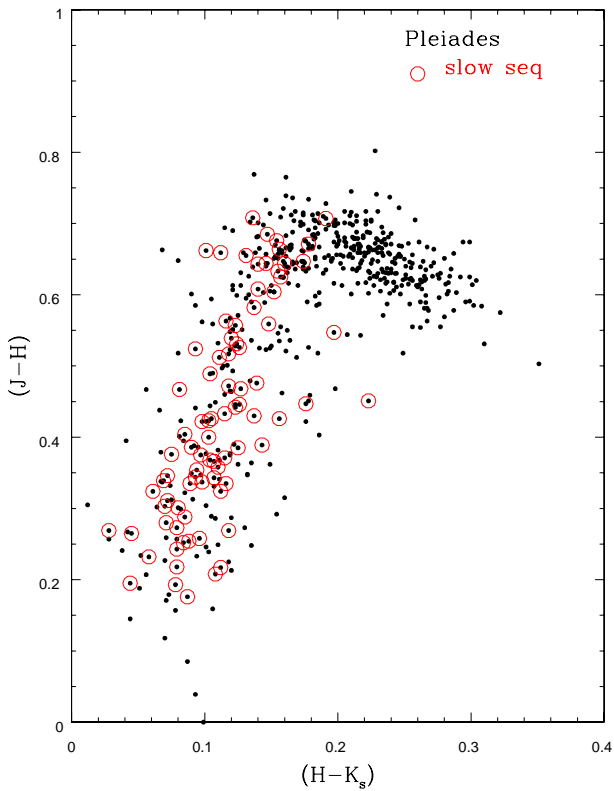}
\caption{$J-H$ vs.\ $H-K_s$ color-color diagram for Pleiades stars
with K2 rotation periods, highlighting members of the slowly rotating
sequence. The coolest slow sequence stars fall just at the mass where
$J-H$ stops getting redder to lower mass, signaling the reduction in
H$^-$\ opacity and the ramp up of molecule opacities for lower masses.
\label{fig:Figure10}
}
\end{figure}

While Figure~\ref{fig:Figure9} is quite persuasive, it is 
nevertheless true that the red edge in the Pleiades occurs almost
precisely at the boundary between spectral types K and M.   That
boundary corresponds approximately to the point where molecules become
prevalent in the photospheres of low mass main-sequence stars, and
hence at the point where free electrons become less prevalent (Mould
1976).  This point is illustrated in a $J-H$ vs.\ $H-K_s$ color-color
diagram on the K2 Pleiades stars (Figure~\ref{fig:Figure10}), where we
have highlighted the stars in the slow sequence. It is apparent that
the slow sequence in the Pleiades ends almost exactly where the dwarf
sequence colors turn over in the $J-H$ vs.\ $H-K_s$ diagram (i.e.,
where $J-H$ stops getting redder as \teff\ decreases and instead
slowly becomes bluer with decreasing \teff).   It could be that it is
simply coincidental that the red limit to the slow sequence happens to
occur at Pleiades age at exactly this color.  Or, it may be that there
is a (possibly small) decrease in the average angular momentum loss
rate at this color as a result of the reduced ionization fraction in
the photosphere, such that the red limit to the slow sequence stalls
at this color for a time rather than increasing more or less linearly
with age.   We will return to this topic at the end of the next section.

\subsection{Is the Kink in the Slow Sequence at V-K $\sim$ 2.6 Real?  
  Is It Present in Other Young Clusters? }

\begin{figure}[ht]
\centering
\includegraphics[width=9cm]{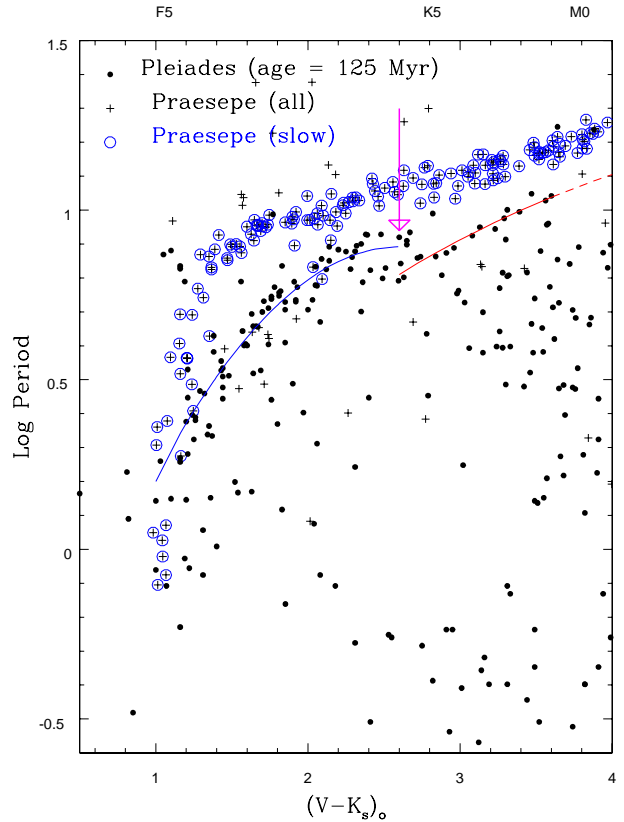}
\caption{Blow-up of the $\log P$ vs.\ \vmk\ diagram for the Pleiades
and Praesepe, highlighting the region of color-period space inhabited
by the slowly rotating sequence of stars in the Pleiades.  The plot
illustrates that the Pleiades slow sequence (a) overlaps with the
Praesepe slow sequence at about \vmk\ = 1.1, and hence that color
marks where significant main sequence angular momentum loss from winds
begins; (b) the two sequences have maximum separation amongst the G
dwarfs; (c) the separation between the two sequences reaches a minimum
at about \vmk\ = 2.3; (d) there is a  discontinuity (a kink) in the
Pleiades slow sequence at \vmk\ $\sim$ 2.6; and (e) the Pleiades slow
sequence ends at \vmk\ $\sim$\ 3.7 and P = 11.0 days (versus P=15.0 days in
Praesepe at that color).  The magenta arrow marks the location of the
kink in the Pleiades slow sequence. 
\label{fig:Figure11}
}
\end{figure}

We noted a kink at \vmk\ $\sim$ 2.6 in the slowly rotating Pleiades
sequence  in \S 2.   We highlight that kink here in
Figure~\ref{fig:Figure11} with an  expanded plot of just the relevant
portion of the $\log P$ vs.\ \vmk\ diagram, which again compares stars in
the Pleiades and Praesepe.  The curves are the second order polynomial
fits to the Pleiades slow sequence stars described in \S 2.  There is
obviously no kink in the Praesepe slow sequence near \vmk\ = 2.6, but
the kink seems fairly evident in the Pleiades.\footnote{The kink is
also visible in Figure 12 of Hartman \etal\ 2010 at $M_K\sim$ 4.7,
using ground-based periods for the Pleiades.  Our figure and theirs
share many of the same stars, so this does not provide significant
confirmation of the existence of the kink.}  This kink in the
period-color distribution could simply be illusory and attributable to
small number statistics.  However, data exist both in other clusters
and within our K2 Pleiades observations which suggest the kink may be
real; we describe these data now.

\begin{figure}[ht]
\centering
\includegraphics[width=9cm]{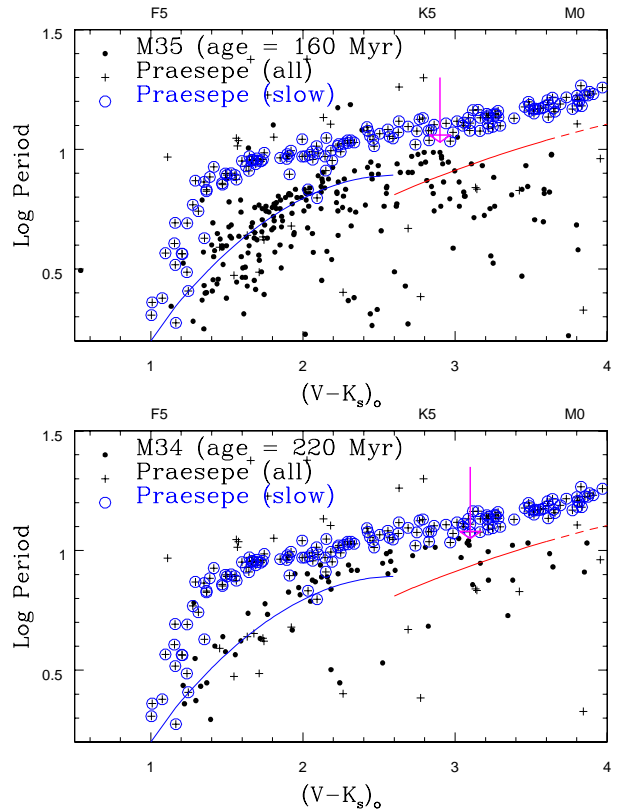}
\caption{Same as for Figure~\ref{fig:Figure11}, except this time 
comparing ground-based period data for 
M35 (age = 160 Myr; Meibom \etal\ 2009) and
M34 (age = 220 Myr; Meibom \etal\ 2011) to
Praesepe. The two curves are the fits to the Pleiades slow sequence
stars.  The M35 periods are
not definitely displaced relative to the Pleiades except near
spectral type K5, where the separation between the M35 and Praesepe
sequences reaches a minimum.  
The M34 figure demonstrates well that M34 has an age
intermediate between that for the Pleiades and Praesepe because the
M34 slowly rotating locus falls above the Pleiades curves and below
the Praesepe stars -- though there is some overlap between stars in
Praesepe and stars in M34 in the K dwarf mass range (illustrating the
stastical nature of ages inferred from rotation).  
There are possible kinks (marked by
magenta arrows) in the M34 and M35 slow sequences at  \vmk\ = 3.1 and
2.9, respectively, similar to the kink in the Pleiades distribution
at \vmk\ = 2.6.  See text for a discussion.
\label{fig:Figure12}
}
\end{figure}

The presence of a kink in the slow sequence at Pleiades age ($\sim$
125 Myr) but not at Praesepe age ($\sim$600 -- 800 Myr) suggests this could
be an age-related phenomenon.   If so, other clusters of similar age
to the Pleiades should  also show the kink.   Large samples of
rotation periods have now been published for many open clusters (\eg\
Irwin \etal\ 2007, 2008. 2009; Messina \etal\ 2010; Barnes \etal\
2015). However, in many cases the exposure times and areal coverage
for those surveys are such that M dwarfs are well-sampled in the data
but relatively few K dwarfs are included, making those data unsuitable
for our purposes.   Two relatively young clusters for which rotation
periods have been published with good coverage of the K dwarf regime
do exist though; these two clusters are M35 (age $\sim$ 160 Myr)
and M34 (age $\sim$ 220 Myr).   Figure~\ref{fig:Figure12} compares the
rotation period distribution for those two clusters to Praesepe's
rotation data, where the rotation data for these  clusters comes from
ground-based campaigns reported in Meibom \etal\ (2009) and Meibom
\etal\ (2011).   Close examination of those plots shows:
\begin{itemize}
\item The M35 slow sequence is very similar to the Pleiades slow
sequence (as previously noted by Hartman \etal\ 2010).  The main
difference relative to the Pleiades is that the feature which could be
interpreted as a kink occurs at a color significantly redder than for
the Pleiades, at \vmk\ $\sim$ 2.9.
\item The M34 slow sequence is exactly as expected, with a locus that
is between that of the younger Pleiades and older Praesepe.   Its
slow sequence has what may be a similar kink to the Pleiades, but at a
redder color than either the Pleiades or M34 (\vmk\ $\sim$ 3.1), 
with a few stars redder than
that which loosely follow the lower branch of the Pleiades slow locus curve.
\end{itemize}

\begin{figure}[ht]
\centering
\includegraphics[width=9cm]{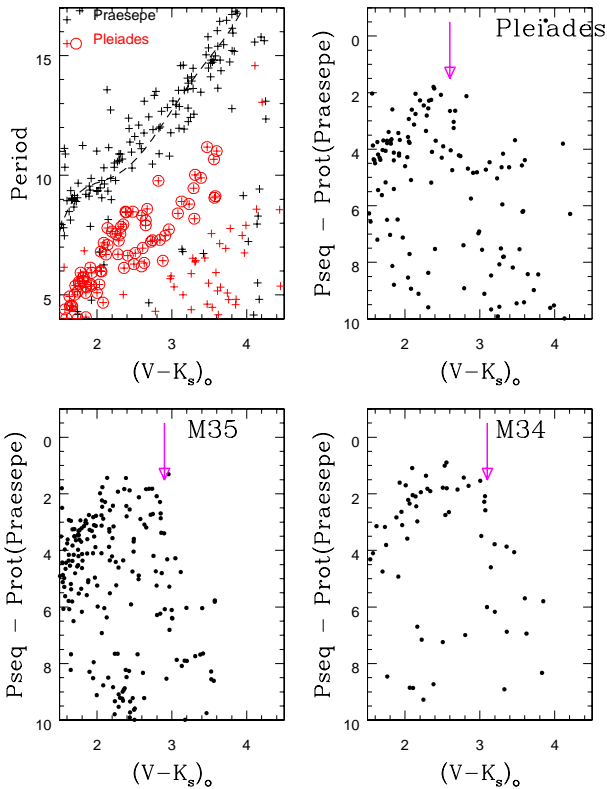}
\caption{(top left) Comparison of rotation periods in the Pleiades and
Praesepe, showing just the more slowly rotation FGK stars, with
the y-axis now being linear, with period in days.  The stars in
the slow rotator sequence of the Pleiades are circled in red; 
(top right) the same Pleiades
data, also with a linear y-axis, but this time plotting the difference
between the observed period and the period for a star of that color
having the nominal slow-sequence period in Praesepe; (bottom left) same 
as for the previous plot, except for M35; (bottom right) same as for 
the previous plot, except for M34.
The magenta arrows mark the \vmk\ color of the ``kink" discussed
in the text; the color of the kink gets redder in step
with the age of the cluster
\label{fig:Figure13}
}
\end{figure}

A logarithmic scale can sometimes be misleading and can under-emphasize
real but small effects.   In Figure~\ref{fig:Figure13}, we replot a
portion of the period data for the young clusters with a linear y axis.
The top, left panel of Figure~\ref{fig:Figure13} shows the Pleiades and Praesepe stars, 
zooming in on the region around \vmk\ = 2.6.   For the other three panels 
of Figure~\ref{fig:Figure13}, for each cluster 
we plot the difference in period between
a star's observed period and the period appropriate to that color for
a star in Praesepe (using the polynomial fit to the Praesepe period
data shown in the first panel).  These plots emphasize both the reality of
a change in the period distribution at the colors we have advocated
in the previous paragraphs and the fact that the location of this
effect progresses to redder colors for older clusters.

What is somewhat concerning is that the red branch of the
Pleiades slow sequence (the stars with red-circles and \vmk\ $>$ 2.9
in the first panel, or correspondingly the stars with $\Delta$P $\sim$ 4 in
the 2nd panel) seems much less populated in M34 and M35.  This could be
a selection bias -- these stars would be precisely the ones most
difficult for ground surveys to detect their periodicity because of
their expected long periods, relatively small amplitudes, and likely
strong spot evolution/migration (as found in the Pleiades -- see next
few paragraphs).   Alternatively, the Pleiades may be unusual for a
young cluster in having the red portion of the slow sequence so well
populated.   In either event, this could argue against designating the
red edge  of the slow sequence (\vmk\ = 3.7 for the Pleiades) as an
empirical  signpost of age for young clusters. Instead, the color of
the kink in the distribution (\vmk\ = 2.6, 2.9, and 3.1 for ages of
125, 160 and 220 Myr, respectively) could be a better age indicator.  
M35 was observed by K2 in Field 0 and so it may eventually be possible to use
those data to derive periods for more late K dwarfs in that cluster and
possibly determine better the shape of its slow sequence and thereby
allow a more detailed comparison to the Pleiades rotational data.

\begin{figure}[ht]
\centering
\includegraphics[width=9cm]{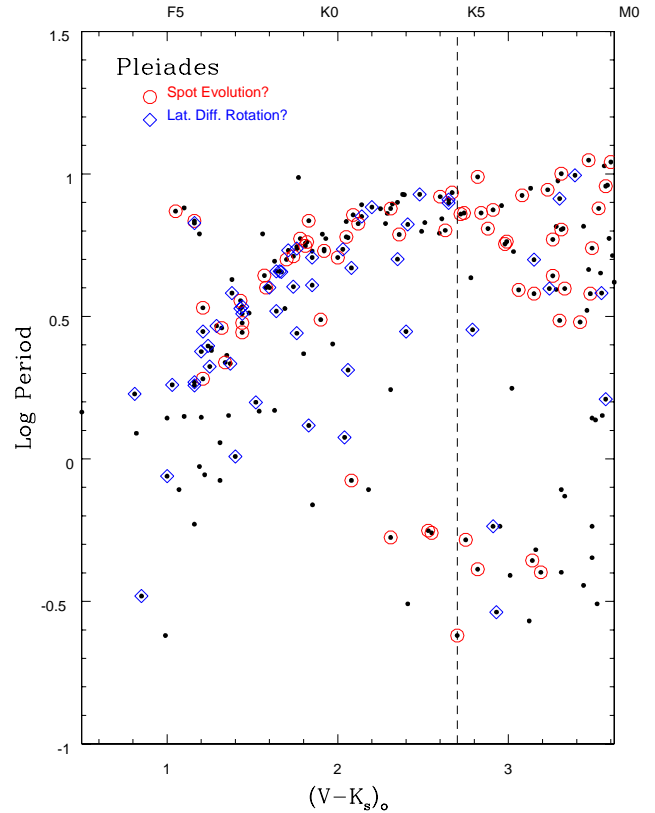}
\caption{Period-color plot for FGK stars, marking stars  
with differing light curve morphologies as deduced in Papers I and II.
Red circles correspond to stars with single-peaked Lomb-Scargle
periodograms but evolving light-curve shape; blue diamonds correspond
to stars with two, close peaks in their Lomb-Scargle periodogram (excluding
``double-dip" stars).  
Bluer than \vmk = 2.6, the former signature 
dominates; redder than that point, the latter signature is significantly more
frequent.
\label{fig:Figure14}
}
\end{figure}

If the kink in the Pleiades slow sequence period-color relation  marks
a change in magnetic field properties or wind properties, it is
reasonable to believe that this change could be reflected in other
measureable properties of the K2 light curves.  As noted in \S2
(based on the histograms shown in Figure~\ref{fig:Figure3}), the
fraction of slowly rotating stars does change abruptly at this color,
going from $\sim$60\% for \vmk $<$ 2.6 to of order 30\% for \vmk
$>$ 2.6.  There is also a marked change in the morphology of the K2
light curves at that color.  Figure~\ref{fig:Figure14} again shows the
period vs.\ color plot for the Pleiades, but this time marking stars
found in \S 3.4 of Paper II to have two resolved, close peaks in their
Lomb-Scargle periodogram suggestive
of either latitudinal differential rotation or spot evolution 
(blue diamonds in the figure) and
those stars with single-peaked Lomb-Scargle periodograms but with
evolving light curve shape (red
circles in the figure). Seventy-eight percent of the FGK stars in the
slow, intermediate and rapidly rotating sequences in
Figure~\ref{fig:Figure2} have one or both of these signatures, versus
a negligible fraction for the M dwarfs.  This alone illustrates that
the magnetic field morphology differs between the GK dwarfs and the M
dwarfs.  Blueward of \vmk = 2.6, most (54\%) of the symbols are blue,
meaning that stars with two close, resolved peaks in their periodograms
predominate;
redward of \vmk = 2.6, most (76\%) of the symbols
are red, corresponding to single-peaked periodograms and
evolving light curve shape.  The driver of the morphology
changes in either case could be spot evolution or latitudinal differential
rotation or a combination of both - here we only note that 
there is an empirical change in the frequency of the two morphologies
at this \vmk\ $\sim$ 2.6.
While these effects are not definitive, they hint at a change in the
average magnetic field topology at this boundary. Because this
boundary appears to move to lower mass with increasing age, it must be
driven by some internal process which depends on those two properties.

\section{The Rotational Periods of M Dwarfs in the Pleiades}

\subsection{Mass-Period Relations for Young M Dwarfs}

Analysis of ground-based rotation period studies of two nearby, young open
clusters (NGC~2547, age $\sim$ 40 Myr, Irwin \etal\ 2008; NGC~2516,
age $\sim$ 150 Myr, Irwin \etal\ 2007) has been used to argue that the rotation
period distribution for ZAMS age M dwarfs follows a $J
\propto M$ (or, equivalently, a $P \propto M^2$) relation. Those
authors note that stars in this same mass range, 0.1 $<$ $M$/\msun $<$
0.4, in star forming regions with ages of a few Myr also follow a $J
\propto M$ relation. 
The Pleiades K2 data provide a larger sample of well-determined
rotation periods in this mass range than are available for any other
cluster;  we use those data here in order to determine how those
periods depend on mass (and how this compares to the stars in NGC~2547
and NGC~2516). We use the Baraffe \etal\ (2015, hereafter BHAC15) 120
Myr isochrone to estimate masses for the Pleiades stars, as described
in Appendix A. We also choose to reexamine how the Pleiades
(ZAMS-age) period-mass relation  evolves with time by comparing it to
period data for NGC~2264 (we describe the NGC~2264 period data and
mass determinations in Appendix B).  We have compiled our own list of
NGC~2264 periods, determined \vmk\ colors for those stars, estimated
masses using the same BHAC15 models as for the Pleiades (except now
for 3 Myr), and culled the sample to remove stars with strong IR
excesses, large reddening or that fall below the main locus of
NGC~2264 members in a color-magnitude diagram).

\begin{figure}[ht]
\centering
\includegraphics[width=9cm]{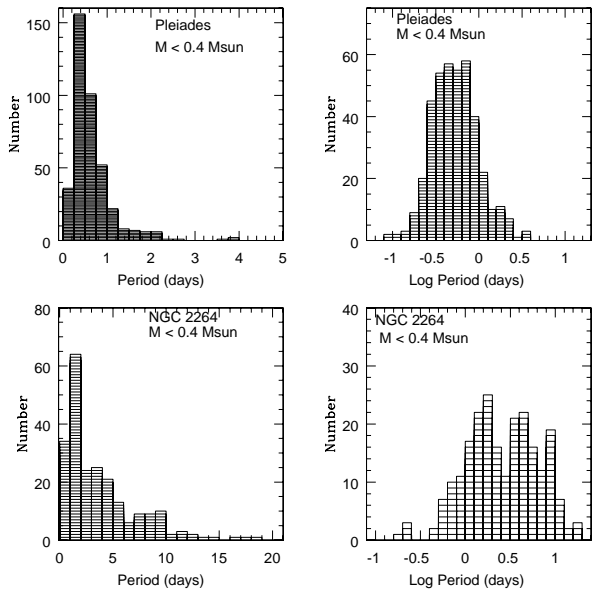}
\caption{Histograms of the rotation period distribution for late-type
M dwarfs in the Pleiades (top) and NGC~2264 (bottom),  both using
linear (left panels) and logarithmic (right panels) binning.  The
shape and width of the rotational distribution for late M dwarfs in
the Pleiades is  similar to that measured for late M dwarfs in the few
Myr old NGC 2264, though the distribution in NGC~2264 appears to be
somewhat broader.
\label{fig:Figure15}
}
\end{figure}

Figure~\ref{fig:Figure15} shows the rotational velocity distribution
of late M dwarfs in the Pleiades and NGC~2264, in histogram format. 
Comparison of those diagrams shows that the Pleiades distribution is
similar in shape, but somewhat narrower, than the distribution in NGC~2264.  
That is, the
distributions are unimodal (as opposed to the bimodal  distribution seen
at higher mass), and the 10\% to 90\% width of the distribution when
plotted in terms of log $P$ is about one dex (but broader for NGC~2264).  
The Pleiades stars
obviously have spun up considerably relative to their younger cousins;
the median rotational period at 0.3 \msun\ in NGC~2264 is about 3.8
days, while it is about 0.68 days in the Pleiades, a change by about a
factor of six assuming the Pleiades and NGC~2264 have the same parent
distribution.   According to the BHAC15
models, the moment of inertia for a 0.3 \msun\ star decreases by about
a factor of 12 between 3 Myr and 125 Myr. PMS
angular momentum loss mechanisms therefore, on average, drain about half of the angular
momentum from 0.3 \msun\ M dwarfs during this time interval.   The
similarity in the shape, but somewhat broader distribution in NGC~2264,
suggests that angular momentum loss
rates scale only weakly with rotation period in this mass and age regime.

\begin{figure}[ht]
\centering
\includegraphics[width=9cm]{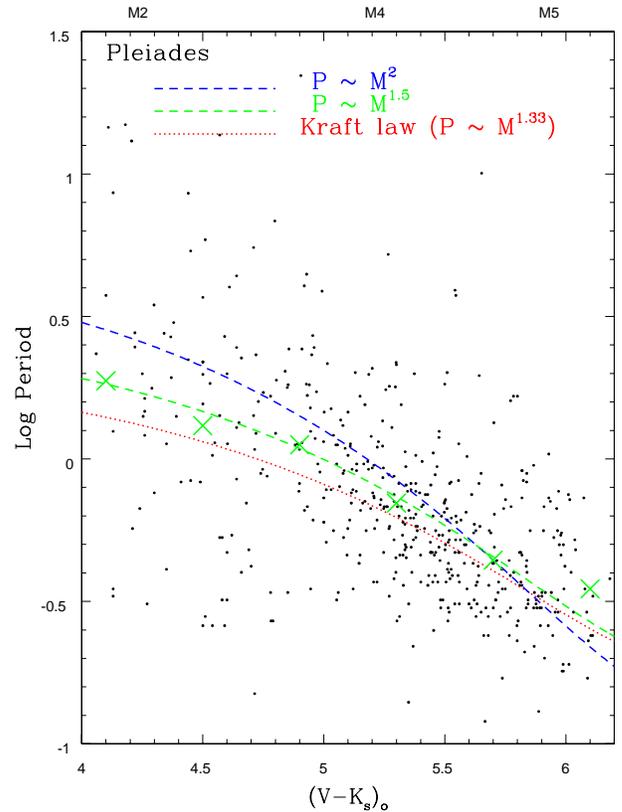}
\caption{Period-color plot for M dwarfs in the Pleiades.  The three
curves correspond to three power laws in mass - with exponents of
1.33, 1.5 and 2.0 for red, green, and blue, respectively.  The $P
\propto M^{1.33}$ relation approximates the Kraft law.   The
green crosses mark median periods in five color bins of 0.4 mag width
beginning at \vmk\ = 4.1.  The mean masses  corresponding to each bin
are 0.56, 0.48, 0.40, 0.30, 0.22 and 0.17 \msun, respectively.  The
best fit is for $P \propto M^{1.5}$.
\label{fig:Figure16}
}
\end{figure}

In order to determine the dependence of period on mass for the
Pleiades M dwarfs, we have divided the stars into six color bins of
0.4 mag width, beginning at \vmk\ = 4.1 (approximately 0.6 \msun) and
ending at \vmk\ = 6.5 (approximately 0.1 \msun), and determined the
median period in each color bin.   Figure~\ref{fig:Figure16} shows
these median periods and several possible period-mass power law
relations.  For stars somewhat higher in mass than the Sun, Kraft
proposed an initial angular momentum distribution of the form $J/M
\propto M^{2/3}$ (Kraft 1970). With $J \propto  MR^2 \Omega$, and
given that at Pleiades age in this mass range it is reasonable to
adopt the approximation that  $R \propto M$, the Kraft law reduces to
$P \propto M^{4/3}$.   For the Pleiades M dwarfs
(Figure~\ref{fig:Figure16}), the Kraft relation seems a bit too
shallow, whereas a $P \propto M^{2}$ relation appears to be too
steep.   The $P\propto M^{1.5}$ (corresponding to $J \propto M^{1.5}$)
relation appears to be  a reasonably good fit.  In Appendix B,
Figure~\ref{fig:FigureB2}  shows a similar plot for NGC~2264.  In this
case, the best fit is for $P \propto M^{1.15}$. Given the
uncertainties in estimating masses at young ages, it is not certain
that these two results are significantly different.  At 3 Myr,  $R$
scales about as $M^{0.5}$ (rather than $R \sim M$ on the ZAMS), so our
period-mass relation at 3 Myr corresponds approximately to $J \propto
M$). These period and angular momentum power law relationships are in
approximate accord with the previous determinations by Irwin \etal\
(2008, 2009). 

\begin{figure}[ht]
\centering
\includegraphics[width=9cm]{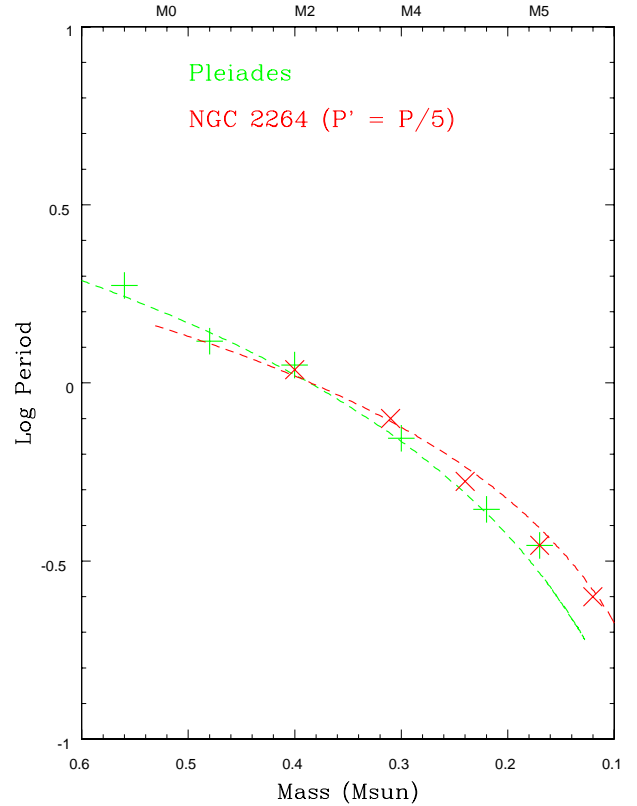}
\caption{
Median periods as a function of mass [as shown in 
Figure~\ref{fig:Figure16} for the Pleiades (green plus signs) and 
and Figure~\ref{fig:FigureB2}) for NGC~2264 (red crosses), 
where the masses have been derived from BHAC15 isochrones
and the periods for NGC~2264 have been divided by five.   The two curves
are the best polynomial fits shown in Figure~\ref{fig:Figure16}
and Figure~\ref{fig:FigureB2}, except again dividing the NGC~2264
periods by five.   When scaled in this fashion, the
median periods for the two clusters are very similar.
\label{fig:Figure17}
}
\end{figure}

To illustrate more directly the similarity of the rotational velocity
distributions for the M dwarfs in the two clusters, Figure~\ref{fig:Figure17}
plots the median periods and best polynomial fits for both NGC~2264
and the Pleiades as a function of mass (as estimated from the BHAC15
isochrones).  In this figure, we have divided all the NGC~2264 periods
by five, in order to approximately account for spinup between 3 Myr
and 125 Myr.  This scaling succeeds remarkably well in aligning the
median periods for the two clusters, indicating that the angular momentum
loss mechanism in this mass and age range preserves the functional
dependence of rotation period on mass. 

The one dex width for the late M rotation distribution in the two
clusters is somewhat misleading because much of that width simply 
reflects the strong correlation between period and mass.   The median
period at \vmk = 5.0 (median mass = 0.39 \msun) is 1.10 days, while at
\vmk = 5.8 (median mass = 0.21 \msun) it is 0.37 days.  To get a
better measure of the true range in $P$ at a given mass for the late M
dwarfs in the Pleiades, we have subtracted an appropriately normalized
power law  ($P = 4.482 \times M^{1.5}$) from the observed periods and
zero-point-shifted the periods to the median period at \vmk=5.0.   The
logarithmic distribution of these normalized periods is shown in
Figure~\ref{fig:Figure17}, where it can be seen that the width of the
distribution is now only about half of that deduced without correction
for the strong correlation of $P$ with mass. 

\begin{figure}[ht]
\centering
\includegraphics[width=9cm]{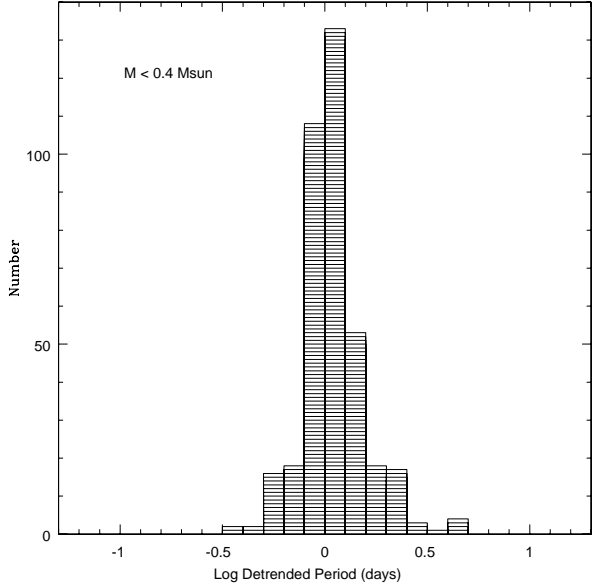}
\caption{Histogram of periods for stars with $M <$ 0.4 \msun in
the Pleiades, after subtracting a power law in mass from the
distribution ($P = 4.482\times M^{1.5})$, and normalizing to the mean
period at $M$ = 0.4 \msun.   The true (mass independent) width of
the distribution is much reduced compared to that shown
in  Figure~\ref{fig:Figure15}.
\label{fig:Figure18}
}
\end{figure}

\subsection{The Initial Angular Momentum Distribution of M Dwarf Binary Stars}

For the Pleiades FGK stars, in Paper II we have associated the presence of close,
resolved peaks ($\Delta$P/P $<$ 0.45) in the LS periodogram with latitudinal differential
rotation or spot evolution on a single star, and distant, resolved peaks ($\Delta$P/P $>$ 0.45) with the
rotation periods of the two stars in a binary.  Theory predicts that
fully convective stars should rotate as solid bodies, therefore we
might have simply assumed that all resolved peaks in the LS
periodogram for mid-to-late M dwarfs were indicative of binary
stars.   However, we prefer to adopt an empirical approach and allow
the data to determine the physical mechanism for multiple peaks in the
LS periodogram of late M dwarfs.   We believe that the Pleiades data
do require a binary star interpretation, but also that those data have
interesting implications for the initial angular momentum distribution
of binary M dwarfs.

For most of the subsequent analysis, we limit ourselves to mid-to-late
M dwarfs (5.0 $<$ \vmk\ $<$ 6.0), with the blue boundary being set at 
where one expects stars to become fully convective and the red
boundary being set where our period data becomes significantly
incomplete due to the faintness of the stars.

\begin{figure}[ht]
\centering
\includegraphics[width=9cm]{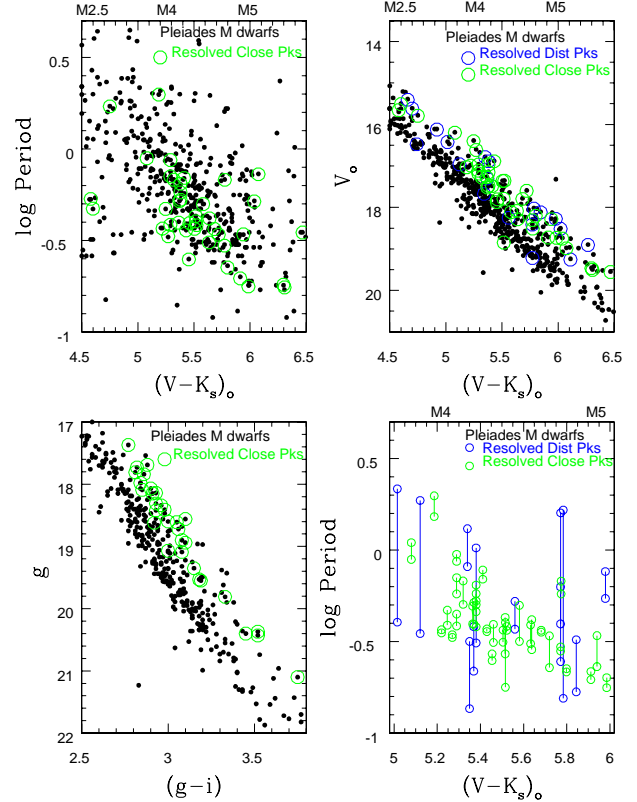}
\caption{(top left) Period-color plot for M dwarfs in the Pleiades,  marking
stars with two close peaks in their Lomb-Scargle periodogram. Note
that the two-close-peaks stars are, on average, relatively rapid
rotators. (top right) CMD for the same set of stars, as well as stars with
distant resolved peaks.  The vast majority of both groups are
displaced well above the single star sequence, suggesting that they
are photometric binaries. (bottom left) A similar CMD, except now for those
stars where we have ugri photometry, showing  qualitatively similar
behavior (and thereby eliminating \vmk\ excesses for the
two-close-peak stars as a plausible explanation for their
displacements in the second panel. (bottom right) A plot showing both periods for each
of the relevant  stars, which emphasizes that in most cases the two
periods are quite close to each other.
\label{fig:Figure19}
}
\end{figure}

The top, left panel of Figure~\ref{fig:Figure19} 
shows where M stars identified in Paper II
as having two close peaks in the LS periodogram fall in a period-color
diagram.  About 10\% of the stars with periods in this color range are
identified with this signature.  The top, right panel of Figure~\ref{fig:Figure19} provides
the $V$ versus \vmk\ diagram for these same stars as well as for stars
identified in Paper II as having distant resolved peaks in the LS
periodogram.   Nearly all of the stars in both categories fall well
above the single-star locus in the CMD, most easily interpreted as
indicating that they are photometric binaries.  This diagram alone is
prima facie evidence that resolved peaks of any form in this color
range should be interpreted as the periods of both components of a
binary because we can think of no reason why latitudinal differential
rotation should only be detectable on photometric binaries. The only
possible escape we can think of for this interpretation would be if
the stars with two resolved peaks were more spotted than single peak
stars and if heavily spotted late M Pleiades members were
significantly redder in \vmk\ than otherwise similar Pleiads. 
The bottom, left panel of 
Figure~\ref{fig:Figure19} shows that this escape route is not viable,
because the resolved close peak stars are just as strongly displaced
above the single-star locus in a $g$ vs.\ $g-i$ CMD, and one expects
$g - i$ to not be significantly affected by spottedness (Stauffer
\etal\ 2003).\footnote{Spots likely do cause Pleiades M dwarfs, particularly
those that are more rapidly rotating, to have redder \vmk\ colors, as
emphasized most recently by Covey \etal\ 2016.  However, this effect
is too small to explain the quite large CMD displacement for the
resolved peaks stars in Figure~\ref{fig:Figure19}.}  
Rappaport \etal\ (2014) also concluded that rapidly
rotating, field M dwarfs with two periods detected in their LS
periodogram are almost always best interpreted as young binary stars.

\begin{figure}[ht]
\centering
\includegraphics[width=9cm]{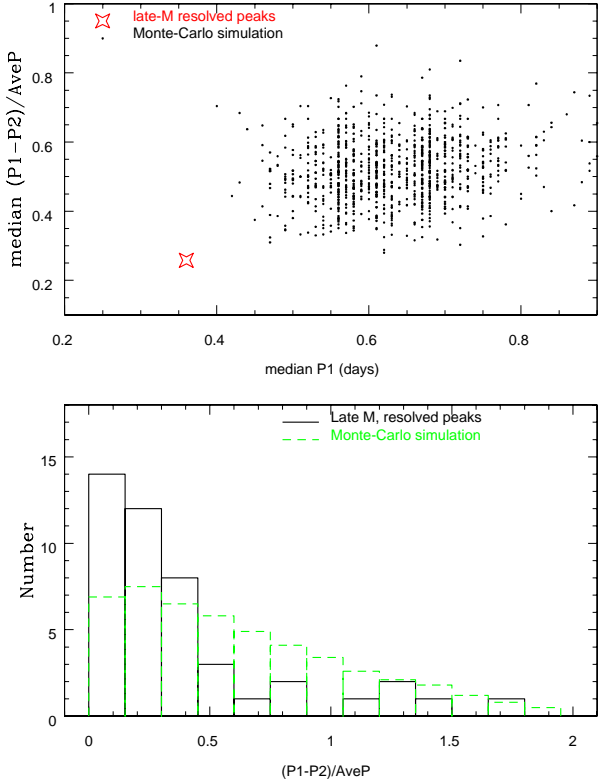}
\caption{(top) Normalized period difference versus median of the first
period for each of the one thousand simulated groups of binaries and
for the observed set of late M Pleiads with two resolved peaks.  The
Pleiades late M stars with two resolved peaks have both shorter 
periods and more closely aligned periods than would have been selected
by chance. (bottom) Histogram of normalized period difference for the
Pleiades late M dwarfs compared to the average for all the simulation
runs, showing a strong excess of small period differences for the Pleiades
binaries.   Nearly 60\% of the late M Pleiades stars with two resolved
peaks in their LS periodogram have period differences less than 30\%.
\label{fig:Figure20}
}
\end{figure}

There is at least one apparent problem with interpreting all of the
resolved peak stars as binaries; this problem is illustrated in
the bottom, right panel of
Figure~\ref{fig:Figure19}, where we now show only fully  convective
stars (5.0 $<$ \vmk\ $<$ 6.0) and we plot both periods for each
star.   Comparing the top, left and bottom, right panels of 
Figure~\ref{fig:Figure19},
it seems that the two identified periods
for a large fraction of the stars with two resolved LS peaks are too
close to each other to have been drawn randomly from the overall
distribution.  We have run a Monte Carlo simulation to test this
assertion, where we created 1000 sets of 45 binaries, with periods
chosen randomly from the observed distribution, selecting stars 
differing in \vmk\ by less than 0.15. Figure~\ref{fig:Figure20}
provides two visualizations of the output from this simulation. 
The top panel of Figure~\ref{fig:Figure20} compares the median normalized period
difference to the median first period for the simulations and the
observed population of two resolved peak Pleiades M dwarfs with 5.0
$<$ \vmk\ $<$ 6.0.  None of the simulated groups have normalized
period differences as small as found for the late M, two resolved peak
stars, indicating  P $<$ 0.1\% for a random selection process. 
The stars in the resolved peak group also appear to be relatively
rapid rotators, with their median period being shorter than the
median period of all of the simulation runs.  The
bottom panel of Figure~\ref{fig:Figure20} compares a histogram of the
observed normalized period differences to that from the average of the
simulated groups.   There is a large excess of systems with period
differences less  than 30\% of the mean period.

A hypothesis which would seem to fit our data is that at the
end of the PMS accretion phase, very low mass (VLM) binaries composed of comparable
mass stars  have periods that are shorter and more similar to each
other than they would be if drawn randomly from all the stars of
that mass in their birth group.  Because there is little dependence
of angular momentum loss rate on either mass or period from that
point to the age 
of the Pleiades (see previous section), this rotation bias is still
retained at Pleiades age.  At much older ages, where mass or period
dependent loss rates become important, the similarity of rotation
periods may become less apparent -- thereby explaining why there has
been little evidence for this effect up to now (though from a
spectroscopic study of 11 field VLM binary stars, Konopacky \etal\
2012 concluded that ``the components of VLM binaries are not randomly
paired in \vsini").  The tendency for late M photometric binary stars to
rotate faster than single stars of the same color is in the same
sense as we found for G and K dwarf binaries in \S 3.1.

\subsection{Inferences for Magnetic Fields and Spot Properties of Pleiades M Dwarfs}

There are suggestive correlations between the   distribution of
rotation periods in the Pleiades and the
measurements of surface magnetic field topologies being made in recent
years with Zeeman Doppler Imaging (ZDI) and other techniques (Donati
\& Landstreet 2009; Linsky \& Scholler 2015).  In particular, those
data show a sharp transition in magnetic field topology at about
spectral type M4, corresponding to about \vmk = 5.0.  Cooler than that
boundary, M dwarfs generally have primarily poloidal, axisymmetric
fields and less complex small scale magnetic structures.  They also
generally show little latitudinal differential rotation.  There is
also a step up in the large scale magnetic energy going from early M
to mid-M (Morin \etal\ 2008, 2010).  Early M dwarfs have strong toroidal
fields and weak poloidal fields, have more small scale magnetic
structure, and often have fairly strong latitudinal differential
rotation.   They also have more rapid evolution in their magnetic
topology, perhaps due to the stronger shearing induced by the
differential rotation (Donati \etal\ 2008).   The sharp transition in
magnetic properties  plausibly occurs at or near the mass where M
dwarfs become fully convective ($M \sim$ 0.35 \msun; spectral type
$\sim$ M3.5).

\begin{figure}[ht]
\centering
\includegraphics[width=9cm]{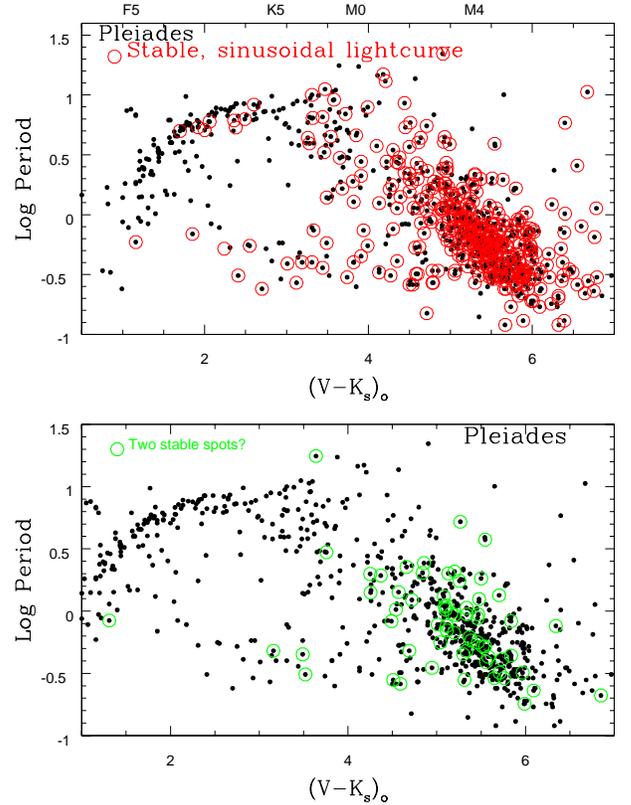}
\caption{(top) Period-color plot for the Pleiades,  marking stars with
with single, narrow peaks in their periodogram, and unevolving,
normally sinusoidal, light curves.   The vast majority of the stars
with this signature are M dwarfs. (bottom) Period-color plot for the 
Pleiades, this time marking stars with two peaks or two minima in
their phased light curves, best interpreted as having large spot
groups in opposite hemispheres of the star. These two light curve
types are the most common morphologies for the rapidly rotating late-M
dwarfs at Pleiades age.
\label{fig:Figure21}
}
\end{figure}

We have already shown in Figure~\ref{fig:Figure14} that the K2 data
corroborate that the GK dwarfs very often show signatures of
spot evolution or
latitudinal differential rotation.   There are other morphological
features of the K2 light curves that may shed light on the magnetic
topologies of Pleiades age M dwarfs.  The top panel of Figure~\ref{fig:Figure21}
highlights a different facet of light curve morphology; the stars
circled in red in this figure have LS periodograms with just one,
narrow peak, and phased-light curves that show no significant
evolution with time and which have sinusoidal or nearly sinusoidal
shapes.  Light curves of this type are almost absent in the slowly
rotating FGK sequence (10\% frequency, for 1.1 $<$ \vmk\ $<$ 3.7), 
whereas among the rapid
rotators in the same color range (the blue circled points in
Figure~\ref{fig:Figure2}), 45\% have light curves of this type.
Amongst the early M dwarfs with 3.7 $<$ \vmk\ $<$ 4.6, 53\%
(48/91) have this signature, whereas for later M dwarfs
with 4.6 $<$ \vmk\ $<$ 6.0, the fraction rises to 73\% (300/410).
The latter number should be regarded as a lower limit because some of
the late M dwarfs are very faint and/or have very low amplitude
variability, making it difficult to characterize the morphological
status of the light curve.   These stable, sinusoidal light curves
imply long-lived spot groups at relatively high latitudes.

The bottom panel of Figure~\ref{fig:Figure21} highlights a different morphological
signature, stars with two or more peaks or minima in their phased
light curves and where there is little or no evolution in the light
curve shape over the time  period of the full K2
campaign.\footnote{The top panel of Figure~\ref{fig:Figure21} is similar to the top,
left panel of Figure 12 of Paper II.  However, the latter figure does
not restrict the highlighted stars to being essentially stable over
the campaign duration nor to having approximately sinusoidal phased
light curve shapes.   Similarly, the bottom panel of Figure~\ref{fig:Figure21} is similar
to the lower, right panel of Figure 13 in Paper II.   Here we show only
the stationary ``double-dip" stars.} That signature is best explained
as due to the presence of spot groups on opposite hemispheres of the
star (see, for example, Figure 5 of McQuillan, Aigrain \& Mazeh 2013
or Figure 4 of Davenport \etal\ 2015), with
little or no differential rotation (and little or no change in spot
size or shape).  

Our K2 data fit well into the magnetic field topology story line for M
dwarfs: 
\begin{itemize}
\item There is a striking transition in the period distribution between
the early and mid-to-late M dwarfs, at apparently the same place where
the B field topologies change.  The early M dwarfs show a broad spread
in  rotation, showing neither a well-defined slow sequence (as present
for the FGK dwarfs) nor a well-defined rapidly rotating sequence (as
present for the late M dwarfs).
\item A very large fraction of the late
M dwarfs show primarily short periods and stable light
curves suggestive of solid-body rotation, long active region
lifetimes,  and primarily axisymmetric,
poloidal fields.  Evolving light curve shapes due to either spot
evolution or latitudinal differential rotation are present for about
a quarter of the early M dwarfs (see Figure 13 of Paper II),
in agreement with the ZDI measurement
of rapid evolution in magnetic topology for some early M dwarfs.
\end{itemize}

\section{Stars of Special Interest}

\begin{figure}[ht]
\centering
\includegraphics[width=9cm]{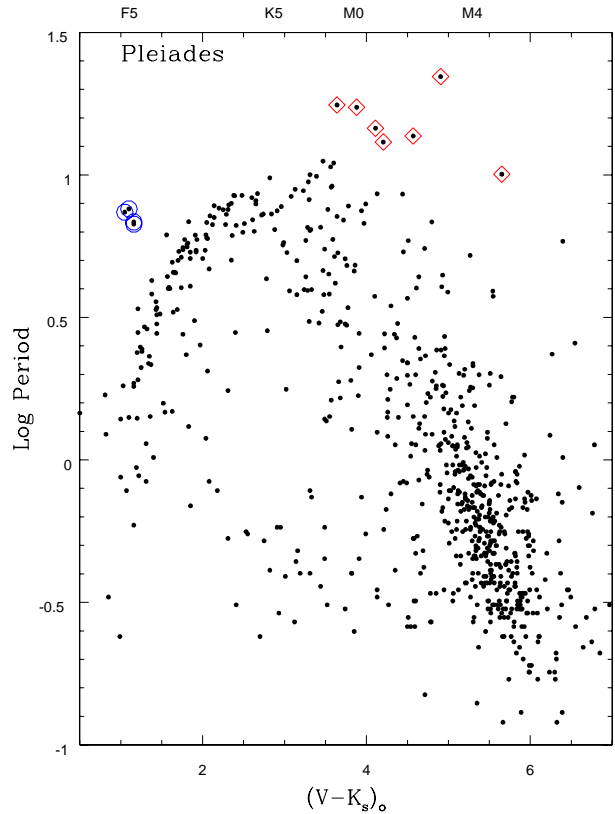}
\caption{Period-color plot for the Pleiades, highlighting two groups
of abnormally slowly rotating stars: blue circles mark a set of very
slowly rotating F dwarfs, red diamonds mark a set of very slowly
rotating M dwarfs.   These two sets of stars are discussed in \S 5.1
and \S 5.2, respectively.
\label{fig:Figure22}
}
\end{figure}

It is often useful to examine outliers to distributions because
their properties may provide clues to not only why they are outliers
but also to the physics that sculpts the distribution for the majority
of stars.  There are two groups of stars whose rotation periods seem
abnormally slow compared to the rest of the Pleiades members; these
two groups are highlighted in Figure~\ref{fig:Figure22}.   We discuss
their properties in the next two sections.

\subsection{Four F Dwarfs with Anomalously Long K2 Periods}

Figure~\ref{fig:Figure22} includes four F dwarfs that seem to be ``lost
in space" -- they all have \vmk $\sim$ 1.1, spectral type $\sim$F4 and
periods near 7 days,  as opposed to an expected period of order 2 days
if they were located on the Pleiades slowly rotating sequence. 
Table~\ref{tab:Fstar_data} lists the  properties of these stars.   All
four stars appear to be very abnormally slowly rotating.   Why is this
the case?

We note that the two simplest explanations do not work.  First,
membership in the Pleiades is not an issue; all four stars are
essentially certain members of the cluster.  Second, their measured
properties as listed in  Table~\ref{tab:Fstar_data}, and in particular
their periods, are well-determined.  

For two of these stars, HII 1132 and HII 1766, the most probable
solution to the conundrum is that the star is a binary, and the K2
period measures the rotation period of the secondary star and not the
primary.   In fact, for these stars, the measured period {\bf{cannot}}
be the rotation period for the primary because the $v \sin i$ values
for these stars are much larger than their inferred rotational
velocity  if one adopts the $\sim$7 day period and a radius of order
1.4 \rsun\ appropriate for their spectral type.  Examination of DSS and
2MASS finding charts for all four stars shows no nearby line-of-sight
companion that would likely contaminate the K2 photometry.

For HII 1132 and HII 1766, one can perform a thought experiment.  If the K2
period derives from a companion star that lies on the slowly rotating
FGK sequence polynomial relation, what would its magnitude be in the
Kepler band and what amplitude would its variability have to be in
order to explain the observed light curve amplitude (where the K2 data
is assumed to include a constant signal from the F star and the
variable light curve from the secondary)?  Doing that experiment for
HII 1132, we find \vmk $\sim$ 2.3 and $K_{\rm kep}\sim$ 12.0, and an
inferred photometric amplitude of close to 7\% for the secondary.  
That is a large amplitude for a Pleiades K dwarf, but within the
range found for slow rotators (see Figure 12 in Paper I).
A similar value would be inferred for HII
1766.   This situation only arises in the Pleiades for F star
primaries, where any intrinsic variability of the F star has very low
amplitude (Figure~\ref{fig:Figure8}) and where plausible, slowly
rotating secondary stars a few magnitudes fainter could have
relatively large photometric amplitudes.  HII 1132 and HII 1766 are
both displaced above the single-star $V$ vs.\ \vmi\  main sequence by about 0.3 mag,
making it possible but not certain that they are binaries with relatively
low mass companions.

For HII 1338, instead, we believe there is a different explanation. 
HII 1338 is a known double-line spectroscopic binary, with a measured orbital period of 7.76d
(Raboud \& Mermilliod 1998).  The stars in the system likely have
their rotation periods tidally locked (or nearly so) to the orbital
period, and that is the physical explanation for their slow rotation
rate.

HII 605 is also a known spectroscopic binary (Pearce \& Hill 1975; Liu et al 1991).  The
radial velocity amplitude is at least 40 km~s$^{-1}$ but no orbital
parameters have been measured.  It is possible therefore that it also
has an orbital period short enough to induce tidal locking at the
observed rotation period; alternatively, if the orbital period is
long, then the explanation proposed for HII 1132 and HII 1776 may
apply.

Finally, we note that HII 1132 has by far the largest IR excess of any
known Pleiades member (Rhee \etal\  2008).   It also has a brown dwarf
companion (Rodriguez \etal\ 2012).   While one could invent possible
connections between these observables and the very slow rotation rate
for HII 1132, with only one object in the sample the connection would
be tenuous.

\begin{deluxetable*}{lcccccc}
\tabletypesize{\scriptsize}
\tablecolumns{7}
\tablewidth{0pt}
\tablecaption{F Stars with Abnormally Slow Rotation Periods\label{tab:Fstar_data}}
\tablehead{
\colhead{EPIC}  & \colhead{Name} & \colhead{Spectral Type} &
\colhead{$(B-V)_o$} & \colhead{\vmk} & \colhead{Period} & \colhead{$v \sin
i$}  \\
\colhead{} & \colhead{} &\colhead{} & \colhead{(mag)} &
\colhead{(mag)} & \colhead{(days)} & \colhead{(km s$^{-1}$)} 
}
\startdata
211121734 & HII 605 & F3 & 0.40 & 1.05 & 7.40 & \ldots  \\
210996505 & HII 1132 & F5 & 0.45 & 1.16 & 6.85 & $>$ 40  \\
211072160 & HII 1338 & F3 & 0.42 & 1.10 & 7.60 & \ldots \\
211138217 & HII 1766 & F4 & 0.43 & 1.16 & 6.72 & 22.7 \\
\enddata
\tablecomments{Spectral types from sources Mendoza (1956), except for HII 1132 which
is from Kraft (1967); photometry and periods from Paper I;
$v \sin i$ values from Queloz \etal\ (1998).  We have a Keck HIRES spectrum for
HII 1132, from which we estimate \vsini\ $\sim$ 50 \kms.}
\end{deluxetable*}

\subsection{The Slowest Rotating M Dwarfs in the Pleiades}

Even in a nearby cluster like the Pleiades, membership studies are
imperfect, particularly as one pushes to relatively faint
magnitudes.   As discussed in Paper I, our original list of potential
Pleiades members with K2 light curves included of order 1000 stars. 
We conducted our own literature search for these stars in order to
verify membership, and ended up eliminating about 150 stars as likely
non-members (see \S 2.5 of Paper I).   Many of these stars either had
no periodic signature in the K2 data, or had quite long periods for
their spectral type if they were Pleiades members (see Appendix C of
Paper I).   However, even after that culling, several M dwarfs remain
that seem to have rotation periods that are abnormally long; these are
the stars marked in red with $P >$ 10.0 days in Figure~\ref{fig:Figure22}.
Are these stars simply non-members that have slipped past our vetting
process, or are they stars which have arrived at Pleiades age with
rotation rates that fall outside the norm for their mass?

\begin{figure}[ht]
\centering
\includegraphics[width=9cm]{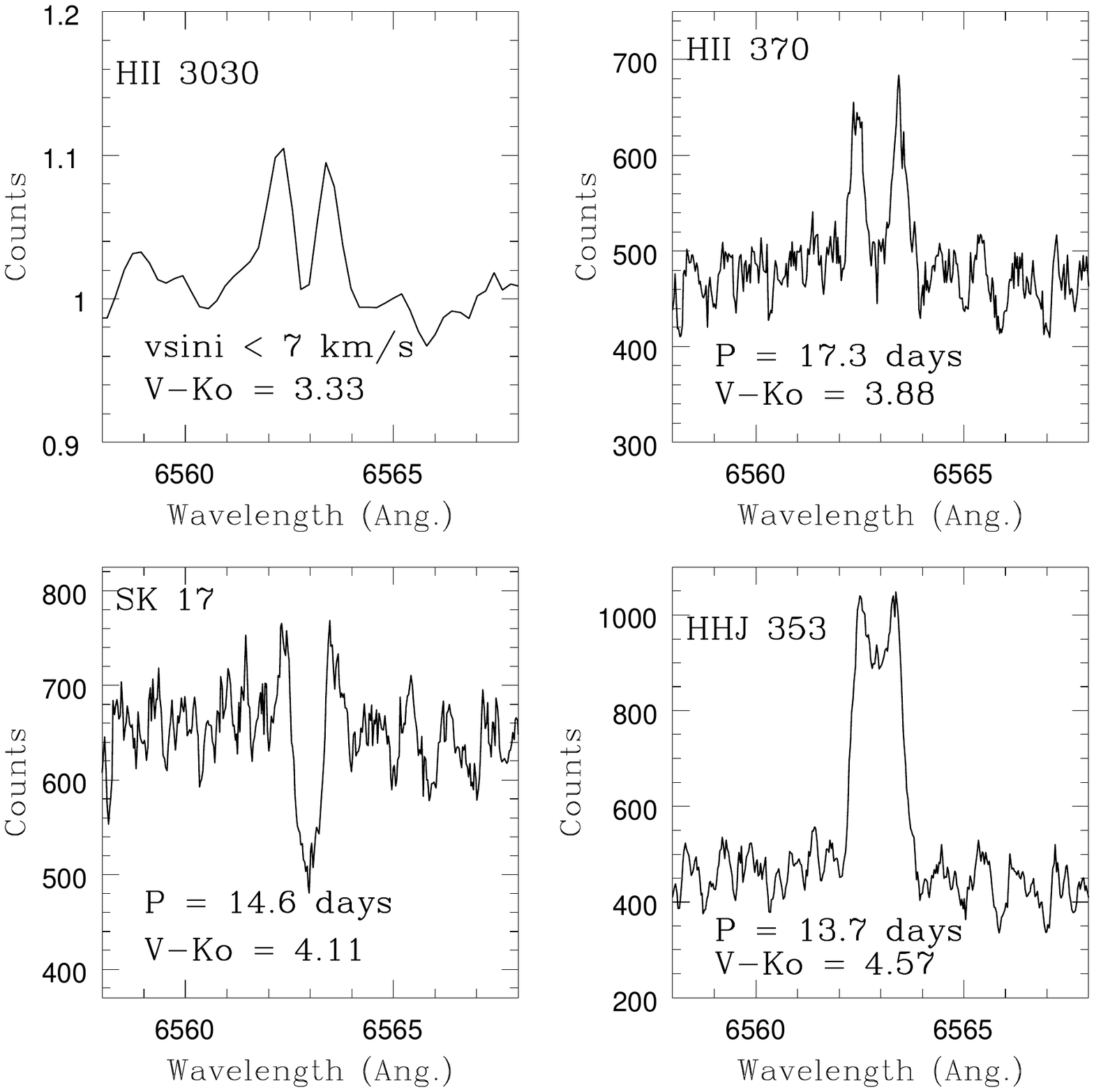}
\caption{High resolution spectra centered on H$\alpha$ for four of the most
slowly rotating late-K or early-M dwarfs in the Pleiades.  The spectrum for HII 3030, from
WIYN/Hydra, is at lower spectral resolution; the other three spectra are
from Keck/HIRES.  
\label{fig:Figure23}
}
\end{figure}

Table~\ref{tab:Mstar_data} provides basic data for these comparatively
slowly rotating M dwarfs, including membership probabilities from  the
most recent survey efforts (Lodieu \etal\ 2012; 
Bouy \etal\ 2015) as well as proper motions from the URAT all sky
survey (Zacharias \etal\ 2015).  All of the stars in this table are
photometric and proper motion members of the cluster based on these
papers and catalogs, and based on our own photometry.   There is
essentially no other published data for these stars that provide
additional useful information on their cluster membership.  To help
interpret their periods, we have obtained Keck HIRES spectra for four
of these stars -- HHJ 353, SK 17, HII 370, and DH 668.   The
membership of DH 668 is uncertain -- our measured radial velocity is
slightly discrepant from the mean Pleiades motion and it has
H$\alpha$\ strongly in absorption, more consistent with it being a
field star.
The other three stars have radial velocities within $\sim$ 1
\kms\ of that expected for Pleiades members.  In addition, all three
have H$\alpha$ equivalent widths suggestive of youth (that is, they
have either weak H$\alpha$ emission or ``filled-in" H$\alpha$).   In 
Figure~\ref{fig:Figure23}, we show the H$\alpha$\ profiles for these
three stars along with a  slowly rotating, late K dwarf Pleiades
member (HII 3030) for which we have a WIYN-Hydra echelle spectrum
(published in Terndrup \etal\ 2000).  We do not have a K2 light curve
for HII 3030; we include it here because of the resemblance of its
H$\alpha$ profile to that for HII 370 and SK 17 and because its 
\vsini\ indicates that it is a relatively slow rotation.l 
HII 3030, HII 370, and SK 17 all
have similar, and slightly unusual, H$\alpha$ emission profiles
consisting of two narrow-emission peaks located in what would have
been the wings of their H$\alpha$ absorption profile if they were not
chromospherically active; HHJ 353 has the strongest H$\alpha$
emission, and has a typical H$\alpha$ profile for a slowly rotating
dMe star.   The H$\alpha$ profiles for HII 3030, HII 370 and SK 17
are essentially as predicted by models for a weakly active dMe star
(Cram \& Mullan 1979). 
The H$\alpha$ equivalent widths for these
four stars are similar to 
or slightly less than other members for their
color.  Given that the three stars from Table~\ref{tab:Mstar_data}
having these new H$\alpha$ data are  more slowly rotating than any
other members of their color, it is probably appropriate that they
also have somewhat weaker H$\alpha$ equivalent widths.   We therefore
conclude that at least these three -- and probably the majority of the
others in Table~\ref{tab:Mstar_data} -- are actual Pleiades members,
and they are indeed rotating abnormally slowly.   Something either in
their initial birth environment or in their angular momentum loss
between birth and Pleiades age has allowed them to spin down more than
all the other stars of their mass.

\begin{deluxetable*}{lcccccccc}
\tabletypesize{\scriptsize}
\tablecolumns{9}
\tablewidth{0pt}
\tablecaption{M Stars with Abnormally Slow Rotation Periods\label{tab:Mstar_data}}
\tablehead{
\colhead{EPIC}  & \colhead{\ks\ } & \colhead{Period} &
\colhead{\vmk\ } & \colhead{Bouy} & \colhead{Lodieu} 
 & \colhead{$\mu$RA}  & \colhead{$\mu$DEC} & \colhead{Name} \\
 & \colhead{(mag)} & \colhead{(d)} & \colhead{(mag)}}
\startdata
211074500 & 13.18 & 10.06 & 5.65 &  0.99      &   0.84       & 27   & -48   &  BPL88  \\
210865020 & 11.24 & 13.05 & 4.21 &  0.88      &   ....       & 16   &  -43  &  UGCSJ03411+205117 \\
211007577 & 11.67 & 13.70 & 4.57 &  0.99      &   0.54       & 23   &  -49  &  HHJ353 \\
211007344 & 11.08 & 14.58 & 4.11 &  0.98      &   0.72       & 16   & -44   &  SK17 \\
211056297 & 10.45 & 17.28 & 3.88 &  0.99      &   ....       & 10   & -59   &  HII370 \\
210855272 & 10.94 & 17.60 & 3.64 &  0.72      &   0.11       & 11   & -38   &  DH668 \\
210975876 & 12.53 & 22.14 & 4.91 &  0.54      &   ....       & 13   & -41   &  UGCSJ040550+223553 \\
\enddata
\tablecomments{Bouy and Lodieu columns are membership probabilities
from Bouy \etal\ (2015) and Lodieu \etal\ (2012), respectively.  The
columns providing proper motions in RA ($\mu$RA) and DEC ($\mu$DEC)
are in units of mas yr$^{-1}$, as provided by URAT proper motion survey
(Zacharias \etal\ 2015).  The mean cluster motion is $\mu$RA = 19.5
mas/year and $\mu$DEC = $-$45.5 mas yr$^{-1}$  (Lodieu \etal\ 2012).
}
\end{deluxetable*}

\section{Summary and Conclusions}

\begin{figure}[ht]
\centering
\includegraphics[width=9cm]{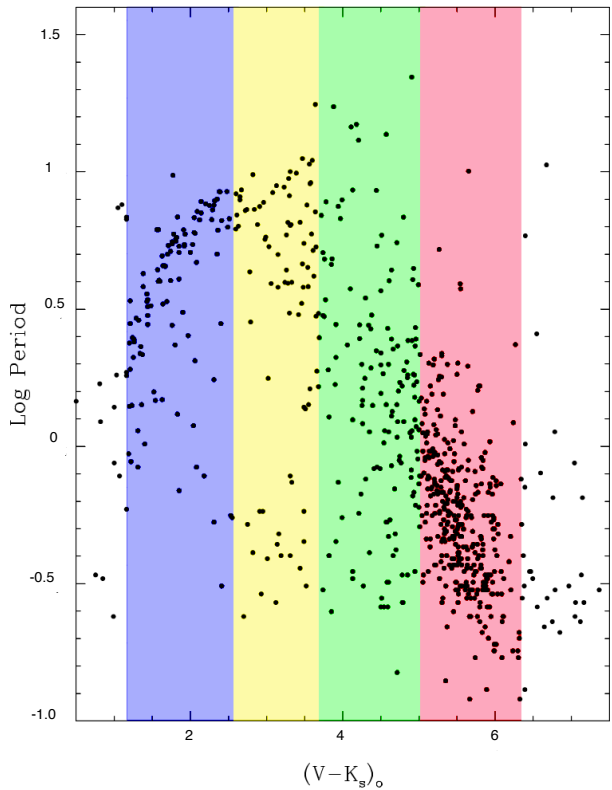}
\caption{Color coding of the Pleiades period vs.\ color diagram,
highlighting the four mass ranges where stars seem to share similar
light curve morphologies; see text.
\label{fig:Figure24}
}
\end{figure}

By providing high cadence, high precision, long duration, sensitive, and
photometrically stable light curves, the K2 data for the Pleiades has
allowed us to derive not only rotation periods and amplitudes, but
also detailed information from the shape of the light curve not
obtainable from ground-based surveys.  In order to help frame the
discussion of the inferences we have made from the K2 data, we provide in
Figure~\ref{fig:Figure24} one last rendition of the Pleiades rotation
periods, this time using changes in the background color to highlight
color (or mass) boundaries where we suspect there are significant
changes in the magnetic field structure or angular momentum loss
rates.  As well known from previous ground-based spectroscopic and
photometric surveys, the G and K dwarfs in the Pleiades show a bimodal
rotation distribution, with large samples of both slowly rotating ($P
>$ 2 day) and rapidly rotating ($P <$ 0.5 day) stars.   The K2 data
additionally show that:
\begin{itemize}
\item The slowly rotating sequence in the Pleiades  extends across the
regions highlighted in blue and yellow in  Figure~\ref{fig:Figure24},
extending from \vmk =  1.1, where $< P >$ = 2.0 days to \vmk\ = 3.7,
where $< P >$ = 11 days, where the sequence appears to end.
\item  Light curve amplitudes increase sharply redward of \vmk = 1.1
(spectral type F5), which we identify as the color where magnetic
dynamos begin to operate efficiently, in agreement with previous
conclusions based on spectroscopic activity indicators and X-rays. 
This is also the color where the Pleiades and Praesepe (age $\sim$ 600 - 800
Myr) slow sequences overlap, indicating that wind-driven angular
momentum loss is ineffective blueward of this boundary.
\item The background color shifts from blue to yellow in
Figure~\ref{fig:Figure24}  to mark a kink in the slowly rotating
sequence, where the mean rotation period suddenly decreases by about
one day.   The fraction of stars on the slowly rotating sequence at a
given color drops sharply at this boundary, from about 60\% blueward
of the boundary to about 30\% redward of the boundary.  
\item The much older Praesepe cluster  shows a slow sequence displaced
throughout the FGK spectral range to slower periods than for the Pleiades and 
extending much further to the red. Praesepe shows no kink in its slow sequence
near \vmk\ = 3.    Plots of the period-color data for the
intermediate age clusters M34 (age $\sim$ 220 Myr) and M35 (age $\sim$
160 Myr) do show possible kinks in their slow sequences at redder
colors than in the Pleiades.  The location of the kink in the slow
sequence may be a better age indicator than the terminal color of that
sequence.
\item Most of the stars on the slowly rotating sequence lie close to
the single star locus in $V$ vs.\ $V-I_{\rm C}$ and $V$ vs.\ \vmk\
CMDs.   Many of the stars with rotation periods just slightly shorter
than the slow sequence stars are photometric binaries.  One possible
explanation for this correlation is that binarity results in shorter
disk lifetimes or disks that are otherwise less able to drain angular
momentum from their host stars.
\item Most of the stars on the slowly rotating sequence also show
evidence of either spot evolution or latitudinal differential rotation
in their light curves.  
\item The portion of Figure~\ref{fig:Figure24} with a red background
highlights mid to late M dwarfs with spectral types roughly M3 to M5.
These very low mass stars have a unimodal period distribution, with
median period decreasing from 1.3 days at \vmk\ = 4.5 to 0.37 days at
\vmk\ = 5.8.   The median M dwarf rotation periods approximately scale
with mass as $P\propto M^{1.5}$, not far from the prediction of the
Kraft law proposed to describe the initial angular momentum
distribution of high mass stars.    
\item Most (73\%) of the mid-to-late M dwarfs have nearly sinusoidal
light curve shapes that change little or not at all in shape over the
duration of the K2 campaign.  Such light curves are best interpreted
as due to a single spot or spot group located at high latitude.  The
next most  frequent light curve type in this mass range are stars best
interpreted as due to two spots or spot groups, displaced in longitude
from each other -- but with little evidence for spot evolution or
differential rotation.   
\item Two close peaks in the LS periodogram are associated with
latitudinal differential rotation or spot evolution for 
FGK stars (see Paper II, Figures 13, 14 and 15), but are instead best
interpreted as the individual periods of components of binary stars
amongst young, late M dwarfs.  We find that the rotation periods of
the Pleiades late M binaries have more similar periods than would be
true if their periods were drawn at random from the overall Pleiades M
dwarf distribution. 
\item The transition region between where the slow sequence ends and
the well-organized, unimodal distribution seen for the mid-to-late M
dwarfs begin is marked with a green background in
Figure~\ref{fig:Figure24}. In this transition region (late-K to 
early-M dwarfs), the rotational
period distribution is less organized than at any other mass.  There
is no well-defined slowly rotating sequence, but the distribution 
still retains some semblance of being bi-modal, with relatively slow rotators ($P
>$ 1 day) predominating over rapid rotators ($P <$ 1 day).
\end{itemize}

Empirical measurements of the magnetic field structures have begun to
illuminate how magnetic geometries change as a function of stellar
mass on the main sequence and at very young ages (Landstreet \& Donati
2009; Gregory \etal\ 2012).  However, as yet there are few empirical
determinations of the magnetic field structures for low mass members
of the Pleiades or for any other young open cluster (see however
Folsom \etal\ 2016 where the first ZDI data on a small sample of
Pleiades stars are reported).   Some of the
light curve morphological properties we find in the Pleiades roughly
match predictions for a solar-type $\alpha$-$\Omega$\ dynamo for the
stars in the blue and yellow regions and to some type of distributed
dynamo (Chabrier \& Kuker 2006) for the fully  convective stars in the
red region.   The green region then represents a transition regime,
where perhaps neither of these two magnetic field generation
mechanisms works well.

Our analysis of the Pleiades K2 data emphasizes that while having
large samples of well-determined periods is important, it is equally
important to also have accurate, homogenous ancillary data in order to
eliminate field star contamination and in order to estimate masses
(\eg, from broad-band colors and extinction corrections)  for all of
the stars with periods.  For the Hyades, Pleiades, Praesepe, and
NGC~2516, the reddening is small enough 
(Taylor 2006; Stauffer \& Hartmann 1987; Terndrup \etal\ 2002)
that whatever uncertainty
exists will not significantly affect any conclusions drawn concerning
the rotation period distributions.   However, for a number of the
other clusters with ground-based data, extinctions are much larger and
their uncertainty can affect the conclusions one reaches.  For
example, the extinction to M35 ($A_V$ = 0.60) adopted by Meibom \etal\
(2009) -- and accepted in our comparisons and those made by Hartman
\etal\ (2010) -- trace back to two unpublished papers; if one were
instead to adopt the best published estimate of $A_V$ = 0.76 (Sung \&
Bessell 1999), the M35 period-color plot (Figure~\ref{fig:Figure13})
would suggest an older age.  The actual extinction is likely to be in
between these two estimates.   To make the best use of the marvelous
precision in period determinations available from space missions and
the best ground campaigns and do precision ``gyrochronology" (Barnes 2010), 
one also
needs to obtain correspondingly good ancillary data.

We view the current paper as but the opening round in a discussion of
not just the Pleiades K2 data, but more generally the use of high
quality light curves as a means to investigate the physics of angular
momentum evolution and magnetic dynamos in young, low mass stars.   We
believe that detailed analysis of such light curves for large samples
of coeval stars can partner well with ZDI and similar direct
measurement of magnetic field topologies to arrive at a better
understanding of these topics than either method could do on its own.

\begin{acknowledgements}
ACC acknowledges support from STFC grant ST/M001296/1.
Some of the data presented in this paper were obtained from the
Mikulski Archive for Space Telescopes (MAST). Support for MAST for
non-HST data is provided by the NASA Office of Space Science via grant
NNX09AF08G and by other grants and contracts. This paper includes data
collected by the Kepler mission. Funding for the Kepler mission is
provided by the NASA Science Mission directorate. 

This research has made use of the NASA/IPAC Infrared Science Archive
(IRSA), which is operated by the Jet Propulsion Laboratory, California
Institute of Technology, under contract with the National Aeronautics
and Space Administration.    This research has made use of NASA's
Astrophysics Data System (ADS) Abstract Service, and of the SIMBAD
database, operated at CDS, Strasbourg, France.  This research has made
use of data products from the Two Micron All-Sky Survey (2MASS), which
is a joint project of the University of Massachusetts and the Infrared
Processing and Analysis Center, funded by the National Aeronautics and
Space Administration and the National Science Foundation. The 2MASS
data are served by the NASA/IPAC Infrared Science Archive, which is
operated by the Jet Propulsion Laboratory, California Institute of
Technology, under contract with the National Aeronautics and Space
Administration. This publication makes use of data products from the
Wide-field Infrared Survey Explorer, which is a joint project of the
University of California, Los Angeles, and the Jet Propulsion
Laboratory/California Institute of Technology, funded by the National
Aeronautics and Space Administration. 
\end{acknowledgements}

{\it{Facility:} \facility{K2}, \facility{Keck}, \facility{Spitzer}.}

\newpage
\newpage

\clearpage

\appendix

\section{Mass Estimates for Pleiades and NGC~2264 stars}

We base the mass estimates in both the Pleiades and NGC~2264 on the
BHAC15 isochrones, adopting ages of 125 Myr and 3 Myr respectively.  
We also adopt distances of 136.2 pc and 760 pc, and A$_v$ values of
0.12 (Pleiades) and 0.00 (NGC~2264).  The latter choice is based, in
part, on our culling the member list to exclude stars with
significant reddening (see Appendix B).

For both clusters, we use the absolute \ks\ magnitude as the primary
input to the mass estimation.   In the Pleiades, we adopt a correction
to the \ks\ magnitude by projecting the observed \ks\ magnitude down
to our adopted single-star \ks\ vs.\ \vmk\ main sequence locus
illustrated in Figure~\ref{fig:Figure4} in order to take binarity into
account.   This effectively means that our mass estimates for binary
stars correspond approximately to the mass of the primary star in the
system.   We then simply adopt the BHAC15 mass appropriate to that
absolute \ks\ magnitude for 125 Myr.   

For NGC~2264, we omit the step to correct for binarity simply because
there is no well defined single-star sequence in its CMD.

Table A1 provides masses, radii, bolometric magnitudes and effective
temperatures for the 853 Pleiades members for which we have 
K2 light curves.   The effective temperature estimates are based on
the Pecaut \& Mamajek (2013) Table 5 data for main sequence stars, using
\bv\ to estimate Teff for the bluest stars and \vmk\ to estimate
\teff\ for
the reddest stars, and a weighted average of the two estimates for
G and K dwarfs.  We
adopted the Pecaut \& Mamajek (2013) bolometric corrections (BC$_V$) to 
estimate $M_{Bol}$, and then the Stefan-Boltzmann law to estimate radii.

\begin{deluxetable}{lccccccccc}
\tabletypesize{\scriptsize}
\tablecolumns{10}
\tablewidth{0pt}
\tablecaption{Derived Data for Pleiades Members with K2 Light Curves\tablenotemark{a}\label{tab:derived}}
\tablehead{
\colhead{Epic}  & \colhead{Name} &
\colhead{\ks} & \colhead{\vmk} &
\colhead{Mass} & \colhead{Mbol} & \colhead{Radius} & \colhead{\teff} & \colhead{$\Delta$V} & \colhead{Comments}  \\
\colhead{} & \colhead{} & \colhead{(mag)} & \colhead{(mag)} &
\colhead{(\msun)}  & \colhead{} & \colhead{(\rsun)} & 
\colhead{(K)} & \colhead{(mag)} & \colhead{} 
 }
\startdata
210754915 & DH343 & 9.41 & 2.31 & 0.85 & 5.67 & 0.90 & 4911 & 0.36 & \\
210762863 & DH318 & 11.04 & 3.94 & 0.59 & 7.80 & 0.60 & 3695 & -0.15 & \\
210769047 & UGCSJ035623.92+192353.3 & 12.88 & 5.71 & 0.22 & 9.80 & 0.35 & 3043 & 0.28 & \\
210770541 & s4543478 & 11.28 & 5.03 & 0.38 & 8.16 & 0.66 & 3241 & 0.72 & \\
210776021 & DH813 & 11.71 & 4.91 & 0.40 & 8.58 & 0.53 & 3284 & 0.16 & \\
210779549 & PELS135 & 8.14 & 1.19 & 1.25 & 3.62 & 1.34 & 6454 & 0.30 & \\
210784223 & s4745026 & 9.97 & 3.12 & 0.73 & 6.53 & 0.84 & 4168 & 0.30 & \\
210784603 & s3289407 & 10.12 & 3.77 & 0.62 & 6.85 & 0.88 & 3785 & 0.62 & \\
210791550 & UGCSJ035916.64+194427.1 & 12.68 & 5.47 & 0.26 & 9.58 & 0.37 & 3106 & 0.15 & \\
210803812 & UGCSJ033910.21+195530.5 & 12.2 & 5.27 & 0.32 & 9.09 & 0.45 & 3164 & 0.19 & \\
210804032 & DH354 & 13.92 & 5.59 & 0.23 & 10.83 & 0.21 & 3073 & -0.84 & \\
\enddata
\tablenotetext{a}{This table is available in its entirety in the online
version. A portion is shown here to demonstrate its form and content.}
\end{deluxetable}

\section{The Rotation Period Distribution for M Dwarfs in NGC~2264}

The Pleiades M dwarf rotational velocity distribution  shows a strong
dependence on mass with a relatively small dispersion at fixed
mass.    Is that distribution primarily a reflection of the initial
angular momentum distribution for stars of this mass, or is it a
reflection of how angular momentum losses between birth and 125 Myr
have shaped the distribution -- or do both factors contribute
significantly to what we see at Pleiades age?    In order to partially
answer that question, we have gathered together all the relevant
rotation data and broad-band photometry for one of the best-studied
star-forming regions -- NGC~2264 -- and analysed those data in
essentially the same way as we have treated the Pleiades data.    We
provide a summary of this process in this appendix; the discussion of
the NGC~2264 rotational velocity distribution and its comparison to
the Pleiades data appear in \S 4 of the paper.

One difficulty with comparing the Pleiades to NGC~2264 is that no one
age can be attached to the stars in NGC~2264; there is instead an
obvious spread in ages (see, for example, Sung \& Bessell 2010), with
the youngest being $<$ 1 Myr and the oldest probably being $\sim$5 Myr
or more.   The distance to NGC~2264 is also not as well established as
one might think, with even recent published values differing by almost
a factor of two. For the present purposes, we simply adopt an age of 3
Myr and a distance of 760 pc (Sung \etal\ 1997) as values that are
representative of the literature.

We compiled rotation periods for members of NGC~2264 from
Rebull \etal\ (2002a), Lamm \etal\ (2004) and Affer \etal\ (2013);
the first two papers derive their rotation periods from ground-based
synoptic photometry, while the 3rd paper used synoptic photometry
obtained with the CoRoT satellite.
We adopted broad-band optical photometry for those stars from 
Rebull \etal\ (2002b) and from Sung \etal\ (2008); we merged
those data with $JH$\ks\ photometry from the 2MASS point
source catalog.

\begin{figure}[ht]
\centering
\includegraphics[width=12cm]{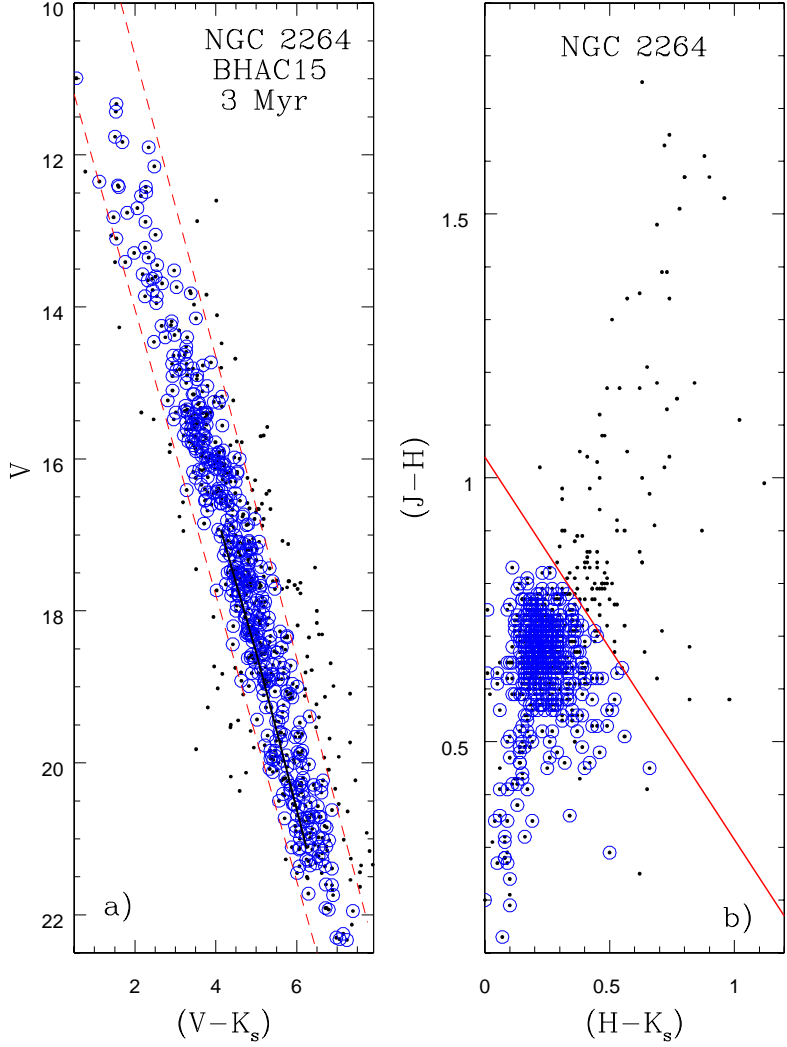}
\caption{(a) $V$ vs. \vmk\ CMD for the stars for which we have
rotation periods in NGC~2264.   The two dashed, red lines delimit the
faint and bright limits within which we retain stars as likely cluster
members with an inferred age near 3 Myr.   The solid black line is the
BHAC15 3 Myr isochrone for 0.5 to 0.1 \msun.  Stars  that survive both
the optical and the near-IR selection criteria are circled in blue. 
(b) The near-IR color-color diagram for the same set of stars.   Stars
above the red line are deleted as either having strong IR exceses
or being heavily reddened.
\label{fig:FigureB1}
}
\end{figure}

Because some of the NGC~2264 stars are classical T Tauri stars (CTTs)
and have IR excesses, and some of the NGC~2264 stars are significantly
reddened due to their being embedded within or behind molecular
clouds, their colors do not necessarily provide an accurate measure of
their photospheric temperature (and hence of their inferred mass).   
Some NGC~2264 members may be much younger or older than average, and
it would be preferable to remove those that are most discrepant (this
step also would preferentially remove field stars that have been
mistaken as members).  We therefore first cull  the sample of stars to
remove outliers,  using an optical CMD and the $J - H$ vs.\ $H - K_s$
color-color diagram.   These two figures are shown in
Figure~\ref{fig:FigureB1}, where we illustrate the cuts in color and
magnitude we have made to purify the sample of stars from which to
determine the 3 Myr rotational velocity distribution.

\begin{figure}[ht]
\centering
\includegraphics[width=9cm]{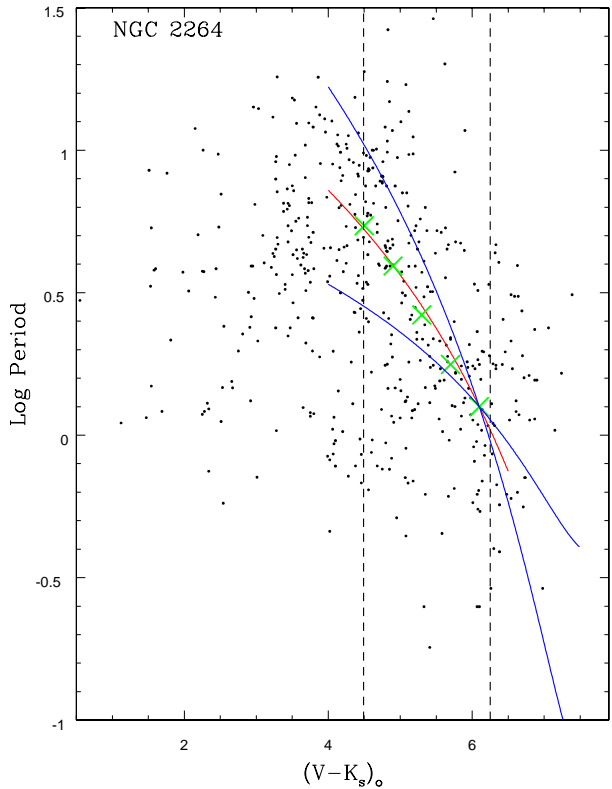}
\caption{Rotation periods as a function of \vmk\ for NGC~2264, for
stars selected to have small reddening, to lack strong IR excesses and
to have colors and magnitudes consistent with the BHAC15 3 Myr
isochrone.  The vertical dashed lines mark the color corresponding to
0.4 and 0.1 \msun; the green crosses are the median periods for all
stars in 0.4 mag color bins beginning at \vmk\ = 4.3-4.7;  the mean
masses for each color bin  are 0.40, 0.31, 0.24, 0.17 and 0.12 \msun,
respectively. The red curve corresponds to a $P \propto M^{1.15}$
power law and is the best fit.  The other two curves are power laws in
mass for exponents of 0.65 and 1.65.
\label{fig:FigureB2}
}
\end{figure}

The period-color diagram for the stars that survived our culling steps
(the stars that are not very young, not very old, not very reddened
and lacking strong IR excesses) is shown in Figure~\ref{fig:FigureB2}.
The vertical dashed lines mark the color corresponding to 0.4 and 0.1
\msun\ according to the BHAC15 3 Myr isochrone.
We have divided the sample into color bins of 0.4 mag width (\eg,
\vmk\ = 4.3 - 4.7, 4.7 - 5.1, 5.1 - 5.5, etc.), and determined the
median rotation period in each bin; those medians are shown as green
crosses in  Figure~\ref{fig:FigureB2}.   Also shown in the figure are
curves for three power-law relations between period and mass --  $P
\propto M$, $P \propto M^{1.15}$ and $P \propto M^{1.5}$ -- all forced
to pass through the median period for the reddest color bin in
NGC~2264.  The $P \propto M^{1.15}$ relation fits the data best.

Finally, we have also constructed histograms of the period distribution
for the 0.1 to 0.4 \msun\ NGC~2264 members, in order to directly compare
to the similar diagrams created from our Pleiades K2 data.  These diagrams
are presented in \S 4.

\section{New Spectra for Very Slowly Rotating Candidate Members of the Pleiades}

In our initial plots of the K2 periods for Pleiades candidate 
members, there were about a dozen early M dwarfs with seemingly
unusually long periods.  In general, these were stars with very little
literature information, and whose membership in the Pleiades was
therefore uncertain.  We obtained high dispersion spectra for eight 
of these stars with  Keck I and HIRES (Vogt \etal\ 1994) on 2-3
February 2016 in order to use those spectra to help ascertain 
membership status.  The instrument was configured to produce spectra
from $\sim$4800 to 9200 \AA\ using the C5 decker which provides 
spectral resolution $\sim$36,000.   We used FXCOR within IRAF to
measure radial velocities, and SPLOT within IRAF to measure the
equivalent widths for H$\alpha$ and to also examine the H$\alpha$\
profile.  We also examined the spectra to check if the \ion{Li}{1} 6708\AA\
doublet was detected; none of the stars showed Li, which was as
expected given their spectral types. Table~\ref{tab:Mstar_HIRES}
provides a summary of the measurements we made for these stars, and
our conclusions concerning membership.  We find that three of the
stars appear to be good Pleiades members, four are most likely
non-members, and one is a possible member.  H$\alpha$\ profiles for
the three probable members are shown in Figure~\ref{fig:Figure22}. 

\clearpage
\begin{deluxetable}{lcccccccl}
\tabletypesize{\scriptsize}
\tablecolumns{9}
\tablewidth{0pt}
\tablecaption{HIRES Spectral Data for Slowly Rotating Candidate Pleiades Members\label{tab:Mstar_HIRES}}
\tablehead{
\colhead{EPIC}  & \colhead{Name} & \colhead{\vmk\ (mag)} &
\colhead{Period (d)} & \colhead{RV (km s$^{-1}$)} & \colhead{H$\alpha$
(\AA)} 
 & \colhead{$\mu$RA}  & \colhead{$\mu$DEC} & \colhead{Membership} }
\startdata
210813780 & DH106    & 4.04 & 15.78 & -25. & +0.31 & 26.5 & -47.4 & probable non-member \\
210855272 & DH668    & 3.64 & 17.60 &   7. & +0.55 & 10.7 & -38.4 & possible member     \\
211182501 & DH836    & 3.32 & 18.59 &  14. & +0.64 & 14.4 & -42.7 & probable non-member \\
211007577 & HHJ353   & 4.57 & 13.70 &   6. & -1.4  & 22.9 & -48.9 & probable member     \\
211056297 & HII370   & 3.88 & 17.28 &   6. & -0.2  &  9.9 & -58.8 & probable member     \\
211094095 & HII813   & 4.55 & 19.15 &  -5. & +0.31 & 31.6 & -48.4 & probable non-member \\
210947847 & s4955064 & 4.19 & 14.89 &  40. & +0.46 & 20.2 & -28.5 & probable non-member \\
211007344 & SK17     & 4.11 & 14.58 &   6. & +0.1  & 15.6 & -43.6 & probable member \\
\enddata
\tablecomments{The columns providing proper motions in RA ($\mu$RA)
and DEC ($\mu$DEC) are in units of mas/year, as provided by URAT
proper motion survey (Zacharias \etal\ 2015).  The mean cluster motion
is $\mu$RA = 19.5 mas yr$^{-1}$ and $\mu$DEC = -45.5 mas yr$^{-1}$  (Lodieu
\etal\ 2012).  The mean cluster radial velocity is about 5.5 \kms.
}
\end{deluxetable}

\end{document}